\documentclass[aps,preprint,floatfix,nofootinbib,superscriptaddress]{revtex4-1}

\pdfoutput=1
\usepackage{graphicx,color,float}
\usepackage{hyperref}
\usepackage{multirow}
\usepackage{verbatim}
\usepackage{longtable}
 \usepackage{amsmath}
\usepackage{amsfonts}
\usepackage{amssymb}
\usepackage{rotating}

\usepackage[normalem]{ulem}

\def\gsim{\raise0.3ex\hbox{$\;>$\kern-0.75em\raise-1.1ex\hbox{$\sim\;$}}}
\def\lsim{\raise0.3ex\hbox{$\;<$\kern-0.75em\raise-1.1ex\hbox{$\sim\;$}}}

\begin{document}

\preprint{APCTP Pre2019-025}

%\title{Probing Long Lived Particles at FCC-ee and CEPC}
%\title{Displaced vertices and Higgs-portal models at the CEPC and FCC-ee}
%\title{Displaced vertices, Higgs-portal, and neutral naturalness at the CEPC and FCC-ee}
\title{Probing Long-lived Particles at Higgs Factories}
\renewcommand{\thefootnote}{\arabic{footnote}}

\author{Kingman Cheung}
\email{cheung@phys.nthu.edu.tw}
\affiliation{Physics Division, National Center for Theoretical Sciences,
	Hsinchu, Taiwan}
\affiliation{Department of Physics, National Tsing Hua University,
	Hsinchu 300, Taiwan}
\affiliation{Division of Quantum Phases and Devices, School of Physics, 
	Konkuk University, Seoul 143-701, Republic of Korea}

\author{Zeren Simon Wang}
\email{zerensimon.wang@apctp.org}
\affiliation{Asia Pacific Center for Theoretical Physics (APCTP) - Headquarters San 31,\\ Hyoja-dong, Nam-gu, Pohang 790-784, Korea}
\affiliation{Physics Division, National Center for Theoretical Sciences,
	Hsinchu, Taiwan}

\date{\today}

\begin{abstract}

  We study displaced vertex signatures of long-lived particles (LLPs)
  from exotic Higgs decays in the context of a Higgs-portal model and
  a neutral-naturalness model at the CEPC and FCC-ee.  Such two models
  feature two representative mass ranges for LLPs, which show very
  different behavior in their decay signatures.  The Higgs-portal
  model contains a very light sub-GeV scalar boson stemming from a
  singlet scalar field appended to the Standard Model.  Such a light
  scalar LLP decays into a pair of muons or pions, giving rise to a
  distinctive signature of collimated muon-jet or pion-jet, thanks to
  the sub-GeV mass.  On the other hand, the neutral-naturalness model,
  e.g., folded supersymmetry, predicts the lightest mirror glueball of
  mass $O(10)$ GeV, giving rise to long decays with a large transverse
  impact parameter because of the relatively large mass.  Utilizing
  such distinct characteristics to remove the background, we estimate
  the sensitivities of searches for light scalar bosons and mirror
  glueballs at the CEPC and FCC-ee.  We find either complementary or
  stronger coverage compared to the previous results in the similar
  contexts.
  
\end{abstract}

\maketitle

\section{Introduction}\label{sec:intro}

Since the discovery of a Standard Model (SM)-like Higgs boson at the
LHC in 2012 \cite{Aad:2012tfa,Chatrchyan:2012xdj}, there has been no
sign of new physics in collider experiments so far.  As most effort at
high-energy colliders e.g. the LHC has been paid for searching for
\textit{promptly decaying} new heavy fields, new physics (NP) that
would emerge in the form of long-lived particles (LLPs) predicted in
various models beyond the Standard Model (BSM) might have been missed,
as the current triggers for events are not designed specifically for
such purposes.  In fact, the rising interest in searches for LLPs has
injected life into both theoretical and experimental communities
\cite{Alimena:2019zri}.  Many BSM models have been proposed that
predict existence of LLPs, either charged or neutral: Higgs-portal
models, $R$-parity violating (RPV) supersymmetry with small RPV
couplings, quirky models, gauge-mediated models, etc. For a summary
see Ref.~\cite{Alimena:2019zri}.  Some searches for LLPs have also
been performed at ATLAS and CMS: see for examples
Refs.~\cite{Aad:2019tua,Aad:2019tcc,Aaboud:2019opc,Aaboud:2018jbr,Sirunyan:2018vlw}.
Nevertheless, the current triggers are not optimized for detecting
LLPs so that perhaps many signals might have been missed.  Specific
triggers will be installed in future runs at the ATLAS and CMS
experiments \cite{Aad:2013txa,CMS:2018qgk}.

Other than the high-luminosity run at the LHC there are also proposals
for the next generation $e^- e^+$ colliders, such as CEPC
\cite{CEPC-SPPCStudyGroup:2015csa}, FCC-ee \cite{Abada:2019zxq}, and
ILC \cite{Djouadi:2007ik}.  These future colliders, if they can be
approved, all involve an important phase of operation as a Higgs
factory running at center-of-mass energies $\sqrt{s} = 240-250 $ GeV.
In such machines, the Higgs bosons are dominantly produced by
Higgsstrahlung together with a relatively small contribution from
$WW/ZZ$ fusion.  Producing the Higgs bosons copiously with relatively
little background, such Higgs factories are expected to be ideal
avenues for precision measurements of the Higgs couplings and other
properties of the Higgs boson, including searching for rare decays of
the Higgs boson.  One particularly interesting rare decay possibility
is the Higgs decay into LLPs.  Though charged LLPs could be easily
identified as stable charged tracks in tracker detectors at the LHC,
neutral LLPs would be easily missed in the current searches.

In this work, we study the sensitivity reach of the CEPC and FCC-ee
for some neutral LLPs produced from rare Higgs decays. We use a
Higgs-portal model and a dark glueball model as the prototype
models. These models share the same features that the neutral LLPs are
scalar bosons pair-produced from the SM Higgs decays, and
followed by their predominant decays into a pair of
leptonic or hadronic jets.  We first investigate the dimuon decay of
the new scalar bosons of the Higgs-portal model. It has the advantage
that the muons can be detected quite cleanly in both the inner tracker
detector (IT) and muon spectrometer (MS).  The detection and
identification efficiencies are in general very high.  The new scalar
boson of the toy models mentioned above has distinct interesting mass
ranges.  The Higgs-portal model \cite{Chang:2016lfq} that we are
interested in allows the new scalar boson as light as sub-GeV.
Because of the light mass, it will be traveling with a high transverse
momentum in the Higgs decay, such that the opening angle between the
dimuons would be order $\Delta R_{\mu\mu} \sim 2 m_{h_s} / p_T =
O(10^{-2})$.  Making use of this feature one can effectively eliminate the
background events. The light scalar boson could as well decay into a
pair of (charged) pions with the same feature as the collimated jets.  We
also take into account this possibility by considering reconstructing
the displaced vertices in the IT, HCAL, or MS.  On the other hand, the
dark glueball lies in the mass range of a few ten's of GeV.  We focus
on the mirror glueball that decays to a pair of $b$-jets, given the fact
that the decay branching ratios of the mirror glueball follow the pattern of the SM-like Higgs boson of the same mass.  Given the relatively large mass of the
mirror glueball, the $b$-jet pair will have a wide opening angle.
These two models thus provide two distinct representatives in the
search of such LLPs. 

The organization of this paper is as follows. In the next section, we
highlight on the two representative models studied in this work. In
Sec.~\ref{sec:simulation}, we describe briefly the layout of the CEPC and
FCC-ee detectors, and detail our search strategies and simulation
procedures. We present the results in Sec.~\ref{sec:results}, and
conclude in Sec.~\ref{sec:conclusions}.

\section{Two Models for the Long-lived Particles}\label{sec:models}

\subsection{A Higgs-portal model}

In this work we consider a toy Higgs-portal model where an additional
real SM-singlet scalar field $X$ is added to the SM Lagrangian, and
the field $X$ mixes with the SM Higgs doublet field $\Phi$, in the
presence of a new $Z_2$ symmetry.  The new scalar field $X$ is odd
under the $Z_2$ such that no $X$ or $X^3$ terms appear, while all the
SM fields are even.  The renormalizable Lagrangian is given by
\begin{eqnarray}
{\cal L} &=& \frac{1}{2}\partial_{\mu}X\partial^{\mu}X
+\frac{1}{2}\mu^{2}_{X}X^{2}-\frac{1}{4}\lambda_{X}X^{4}
-\frac{1}{2}\lambda_{\Phi X}(\Phi^\dagger \Phi)X^{2} \nonumber \\
&+& {\cal L_{SM}} \;, 
\label{L'}
\end{eqnarray}
where the SM Higgs sector is expressed with
\begin{eqnarray}
{\cal L_{SM}} &\supset& (D_{\mu}\Phi)^{\dagger}(D^{\mu}\Phi)
 +\mu^{2}(\Phi^{\dagger}\Phi)-\lambda (\Phi^{\dagger}\Phi)^{2} \;.
\end{eqnarray}
After the electroweak symmetry breaking (EWSB), both the SM Higgs
doublet field $\Phi$ and the new scalar singlet field $X$ are expanded
around their vacuum-expectation values $\langle\phi\rangle\approx 246$
GeV and $\langle\chi\rangle$:
\begin{eqnarray}
\Phi (x) &=& \frac{1}{\sqrt{2}} \left( \begin{array}{c}
           0 \\
           \langle \phi \rangle + \phi(x) \end{array} \right ) \;  \\
 X(x)    &=& \langle\chi\rangle + \chi (x) \;
\end{eqnarray}
We may express the two tadpole conditions by imposing 
$ \partial V/ \partial\phi = 0 $ and $ \partial V/ \partial\chi = 0 $,
with $V$ labeling the scalar potential part of Eq.~(\ref{L'}):
\begin{eqnarray}
\langle\phi\rangle^{2} &=& \frac{4\lambda_{X}\mu^{2}-2\lambda_{\Phi X}
\mu_{X}^{2}}{4\lambda\lambda_{X}-\lambda_{\Phi X}^{2}} \;,\\
\langle\chi\rangle^{2} &=& \frac{4\lambda\mu_{X}^{2}-2\lambda_{\Phi X}
\mu^{2}}{4\lambda\lambda_{X}-\lambda_{\Phi X}^{2}} \;
\end{eqnarray}
Note that if we take the decoupling limit $ \lambda_{\Phi
  X}\rightarrow 0$ from the above equations, we can reproduce the SM
condition of $ \langle\phi\rangle^{2}=\mu^{2}/ \lambda $ and $
\langle\chi\rangle^{2}=\mu_{X}^{2}/ \lambda_{X} $.

One can easily see from Eq.~(\ref{L'}) that the two scalar fields in
the model, i.e. $\phi$ and $\chi$ will mix with each other and form
new mass eigenstates which we label with $h$ and $h_{s}$,
respectively.  The mass terms of the two scalar bosons are
\begin{equation}
{\cal L}_m =  - \frac{1}{2} \left( \phi \; \chi \right )\,
 \left( \begin{array}{cc} 
         2\lambda\langle\phi\rangle^2 & \lambda_{\Phi X}\langle\phi\rangle\langle\chi\rangle  \\
         \lambda_{\Phi X}\langle\phi\rangle\langle\chi\rangle & 2\lambda_{X}\langle\chi\rangle^2 \end{array} \right )\,
  \left( \begin{array}{c} 
           \phi \\
           \chi \end{array} \right ) \;,
\end{equation}
It is possible to rotate $(\phi\; \chi)^T$ to $(h \; h_{s})^T$ 
through an angle $\theta$
\begin{equation}
 \left( \begin{array}{c}
                h \\ 
                h_{s}  \end{array} \right ) 
= 
 \left( \begin{array}{cc} 
            \cos\theta & \sin \theta \\
            - \sin \theta & \cos\theta  \end{array} \right )\,
 \left( \begin{array}{c}
                \phi \\ 
                \chi  \end{array} \right ) 
\end{equation}
The angle $\theta$ has to be small because of various existing
constraints \cite{Chang:2016lfq} so we will focus on small $\theta$
values for the rest of this section. As a result, we may express
the masses of $h$ and $h_s$, the mixing angle $\theta$, and the
interaction term for a 3-point vertex $hh_sh_s$ in terms of the
parameters in Eq.~(\ref{L'}) as
\begin{eqnarray}
m_h^2 & \simeq & 2\lambda\langle\phi\rangle^{2} = (125.10\;{\rm GeV} )^2 \nonumber \\
m_{h_{s}}^2 & \simeq & 2\lambda_{X}\langle\chi\rangle^{2} \nonumber \\
{\cal L}_{hh_{s}h_{s}} &=& -\frac{1}{2}\lambda_{\Phi X}\langle\phi\rangle h h_{s} h_{s} \nonumber \\
\theta & \simeq & \frac{\lambda_{\Phi X}\langle\phi\rangle\langle\chi\rangle}{m_{h}^{2} - m_{h_{s}}^2} \;, \label{eqn:hsparameters}
\end{eqnarray}

Because of its mixing with the Higgs boson, the scalar boson $h_{s}$
can decay into SM particles with the decay rate proportional to
$\sin^2{\theta}$.  We calculate the partial decay widths for $h_{s}
\to \ell^+ \ell^-$ in the following \cite{Gunion:1989we}
\begin{eqnarray}
\Gamma ( h_{s} \to \ell^+ \ell^-) &=& \sin^2{\theta} \, 
  \frac{m_\ell^2 m_{h_{s}}}{ 8 \pi \langle \phi \rangle^2 } \left( 1 - 
  \frac{4 m_\ell^2}{m_{h_{s}}^2 } \right )^{3/2}\;.  \label{eqn:Gammahs2ll}
\end{eqnarray}
For our interested mass range $m_{h_{s}} \alt 1$ GeV\footnote{For slightly higher mass such as 1 GeV $\lsim m_{h_{s}} \lsim 2$ GeV see Ref.~\cite{CidVidal:2019urm} for a search study of such scalars at the LHCb. A sensitivity study for a similar model at the LHC and HL-LHC can also be found in Ref.~\cite{Matsumoto:2018acr}.}, the light scalar
almost only decays to either $\mu^+ \mu^-$, a pair of pions, or four pions, depending on the phase space allowed. For
$h_s\rightarrow \pi \pi$, a similar tree-level analytic expression
given in Ref.~\cite{Gunion:1989we} is insufficient as it fails to take
into account strong final-state interactions near the pion threshold.
Therefore, we adopt the following numerical treatment. We extract
$\Gamma(h_s\rightarrow \pi \pi)$ and $\Gamma(h_s\rightarrow 4\pi)$ from Ref.~\cite{Winkler:2018qyg} and calculate $\Gamma(h_s\rightarrow \mu^- \mu^+)$ with Eq.~\eqref{eqn:Gammahs2ll}, in order to obtain $\Gamma(h_s)$,
the total decay width of $h_s$. Then it is trivial to compute Br$(h_s\rightarrow \pi \pi)$. Further, we calculate Br$(h_{s}\to\pi^{+}\pi^{-})$ with the following formula:
\begin{eqnarray}
\text{Br}(h_{s}\to\pi^{+}\pi^{-}) = \frac{2}{3}
\cdot \text{Br}(h_s\rightarrow \pi\pi),
\end{eqnarray}
since $\Gamma
(h_{s}\to\pi^{+}\pi^{-})=2\, \Gamma(h_{s}\to\pi^{0}\pi^{0})$.
In Table~\ref{tab:BR-hs} we list the decay branching ratios of $h_s$ for $m_{h_s}=0.3-1$ GeV.

\begin{table}[h!]
\caption{\small  \label{tab:BR-hs}
The decay branching ratios for the two most dominant decay modes of the scalar boson $h_{s}$ for $m_{h_{s}} =0.3 - 1$ GeV.
Here $\pi\pi$ includes $\pi^+ \pi^-$ and $\pi^0 \pi^0$.
}
\vspace{1.0mm}
\begin{ruledtabular}
 \begin{tabular}{ l c c c c c c c c }
$ m_{h_{s}}$ (GeV) & 0.3 & 0.4 & 0.5 & 0.6 & 0.7 & 0.8 & 0.9 & 1.0 \\ \hline
$ \text{Br}(\mu^+\mu^-) $ & $ 20.6\% $ & $ 13.0\% $ & $ 10.3\% $ & $ 8.6\% $ & $ 7.1\% $ & $ 5.1\% $ & $ 2.5\% $ & $ 2.0\% $ \\ 
$ \text{Br}(\pi\pi) $ & $ 79.4\% $ & $ 87.0\% $ & $ 89.7\% $ & $ 91.3\% $ & $ 91.2\% $ & $ 93.0\% $ & $ 96.3\% $ & $ 96.8\% $ \\
$ \text{Br}(4\pi) $ & $ 0\% $ & $ 0\% $ & $ 0\% $ & $ 0.1\% $ & $ 1.7\% $ & $ 1.9\% $ & $ 1.2\% $ & $ 1.2\% $ \\ \hline
\end{tabular}
\end{ruledtabular}
\end{table}

\begin{figure}
	\includegraphics[width=0.8\textwidth]{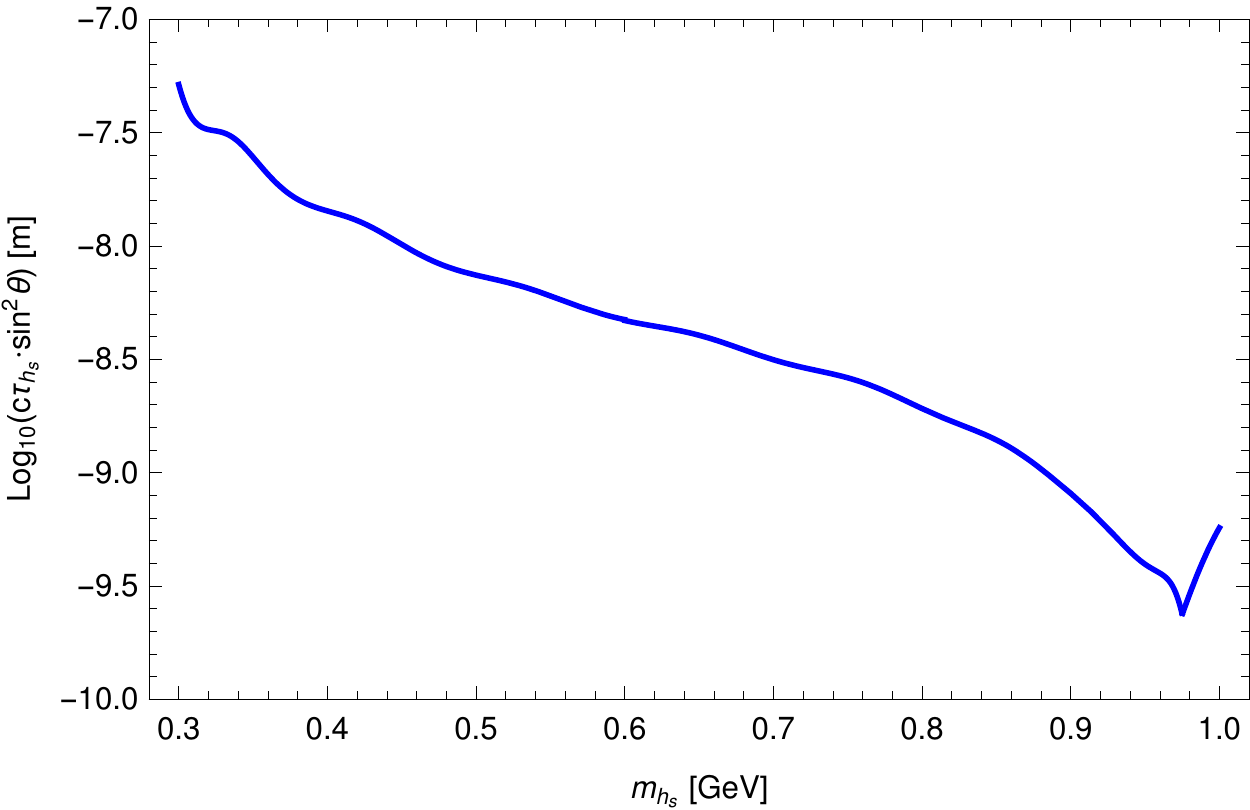}
	\caption{$c\tau_{h_s}\cdot \sin^2{\theta}$ as a function of $m_{h_s}$ in the mass range considered in this work.}
	\label{fig:lifetimehs}
\end{figure}

Before we discuss the production of $h_s$ from the SM Higgs decays, we present a plot of the proper decay length $c\,\tau_{h_s}$ of $h_s$ for $\sin^2{\theta}=1$ as a function of its mass $m_{h_s}$ in Fig.~\ref{fig:lifetimehs}, shown in the plane Log$_{10}(c\,\tau_{h_s} \cdot \sin^2{\theta})$ vs. $m_{h_s}$.

The partial decay width of the Higgs boson into a pair of the light scalar bosons $h_s$ is expressed with the following analytic formula \cite{Chang:2016lfq}:
\begin{eqnarray}
	\Gamma(h\rightarrow h_s h_s) \simeq \frac{\langle\phi\rangle^2}{32\pi m_h}(\lambda_{\Phi X})^2 \simeq \frac{\sin^2{\theta}\, \big( m^2_h - m^2_{h_s} \big)^2}{32  \pi  m_h \, \langle\chi\rangle^2}, \label{eq:GammaH2hh}
\end{eqnarray}
where the first approximation makes use of the fact that the
interested mass range of $h_s$ is negligible compared to the
Higgs-boson mass and so is the phase space effect consequently, and
the second approximation follows from $\sin{\theta}\approx\theta$ for
small $\theta$ and Eq.~\eqref{eqn:hsparameters}.  We can then calculate
the decay branching ratio of the SM Higgs into a pair of $h_s$ as
follows:
\begin{eqnarray}
	\text{Br}(h\rightarrow h_s h_s) =\frac{\Gamma(h\rightarrow h_s h_s)}{\Gamma(h\rightarrow h_s h_s) + \Gamma_h^{\text{SM}}},
\end{eqnarray}
where $\Gamma_h^{\text{SM}}\simeq 4.1$ MeV for $m_h=125.10$ GeV \cite{CERNHiggs}. It is worth mentioning that the invisible decay width measurement of the SM Higgs may constrain $\text{Br}(h\rightarrow h_s h_s)$ and hence the relevant parameter space. Given the current measurement results that $\text{Br}(h\rightarrow \text{invisible})<19\%$ \cite{Sirunyan:2018owy}, we find that $\sin^2{\theta}$ is bounded from above at $\sim 5.1\times 10^{-6}$ and $\sim 5.3\times 10^{-4}$ for $\langle\chi\rangle=10$ GeV and 100 GeV, respectively, and is essentially independent of $m_{h_s}$ since the considered mass range of the light scalar is too small.

Besides the constraint from the current upper bound of the invisible decay of the Higgs boson, searches in $B-$mesons decays at fixed-target, collider experiments, etc., have placed more stringent bounds on the parameter space of a Higgs-portal scalar. However, for the mass range $0.3$ GeV $<m_{h_s}<1.0$ GeV considered in this work, the present limits are completely dominated by the LHCb search results \cite{Aaij:2015tna,Aaij:2016qsm} at $\sin^2{\theta}\sim 10^{-7}$, which searched for $B-$meson decays into a kaon and a light scalar which further decays to a pair of muons.

\subsection{Dark glueball}
  
A class of ``neutral-naturalness'' models are proposed to solve the
little hierarchy problem by predicting the existence of top partners
that are singlet or only charged in the SM electroweak (EW) sector,
which can protect the Higgs-boson mass from large corrections at
one-loop up to some cutoff scales around 5-10 TeV.
Such models of uncolored naturalness usually come with a
dark/mirror QCD sector $SU(3)_\text{B}$, under which the top partner
is charged.  In general, the mirror glueballs lie at the
bottom of the mirror-sector spectrum in these models, including folded
supersymmetry (SUSY) \cite{Burdman:2006tz}, (fraternal) twin Higgs
\cite{Chacko:2005pe,Craig:2015pha}, quirky little Higgs
\cite{Cai:2008au}, and hyperbolic Higgs
\cite{Cohen:2018mgv,Cheng:2018gvu}.
As we will see, the sensitivity 
reach for these models can be derived from one another by a simple
re-scaling, and we therefore focus on one of the models, e.g. the
folded SUSY \cite{Burdman:2006tz}.
In this model, the squarks are charged under $SU(3)_\text{B}$ (but not
the SM $SU(3)_C$ gauge group), and the EW gauge group
$SU(2)_\text{L}\times U(1)_{Y}$ is shared between the SM particles and
superpartners.  Since the LEP limits require that the mirror stops
to be heavier than $\sim 100$ GeV, the mirror glueballs are supposed to be
the lightest states in the mirror sector.  The lightest mirror
glueball $0^{++}$ can be pair-produced from the Higgs boson decay,
followed by the mirror-glueball decay into a pair of SM particles via the top-partner loop-induced mixing with the SM Higgs boson, and so giving rise to displaced-vertex signatures at high energy colliders.

\begin{figure}
	\includegraphics[width=0.8\textwidth]{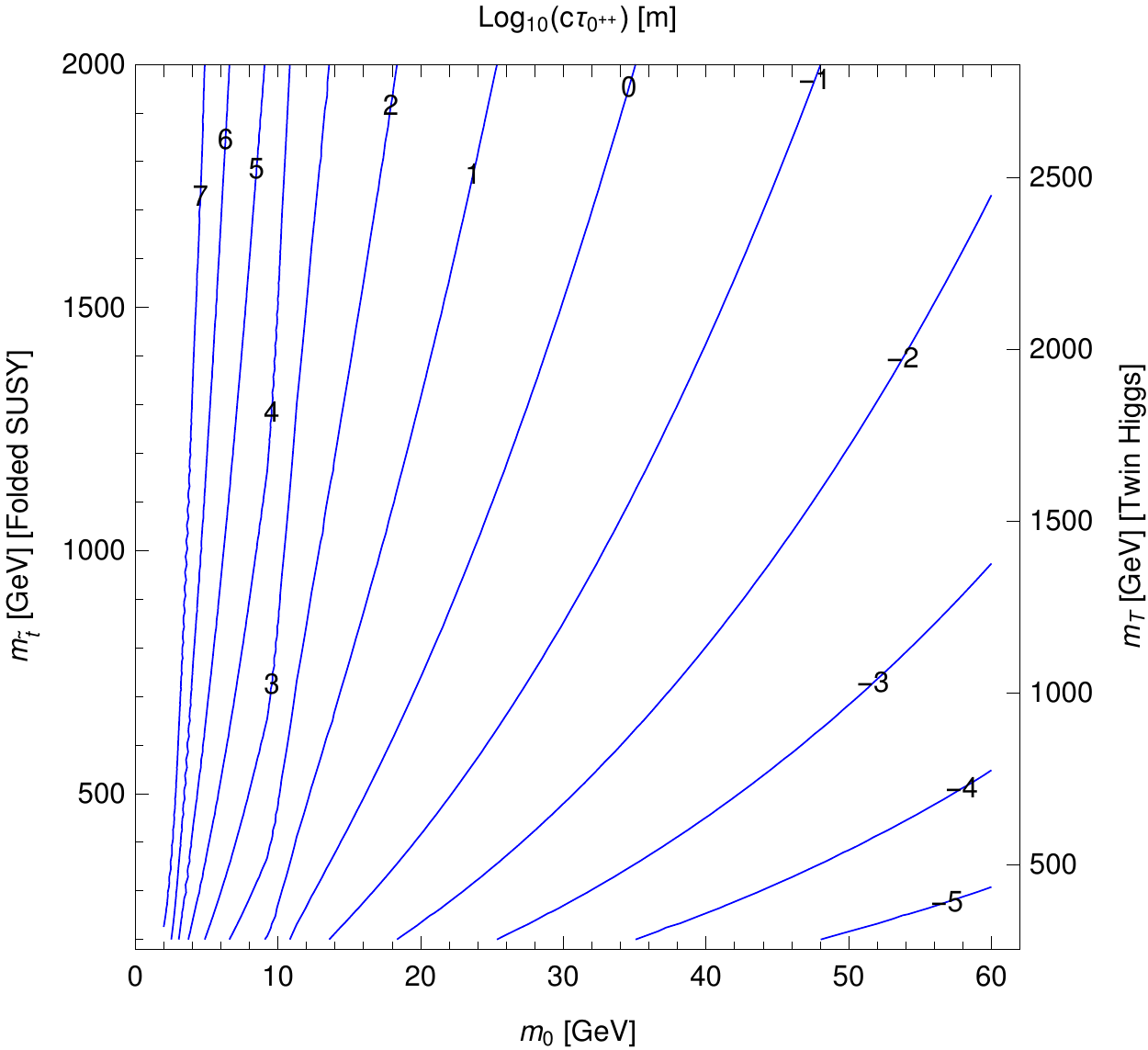}
	\caption{The proper decay length of $0^{++}$ as a function of $m_{\tilde{t}/T}$ and $m_0$.}
	\label{fig:lifetimeglueballs}
\end{figure}

For any scenario of the above mentioned models, the partial decay
width of $0^{++}$ into a pair of SM particles is
given by \cite{Curtin:2015fna,Alipour-Fard:2018lsf,Ahmed:2017psb}:
\begin{eqnarray}
  \Gamma(0^{++}\rightarrow \xi\xi)=\Big(  \frac{1}{12\pi^2}
  \Big[ \frac{y^2}{M^2}\Big] \frac{v}{m_h^2-m^2_0}\Big)^2
  (4\pi \alpha_s^B \mathbf{F^S_{0^{++}}})^2
  \Gamma^{\text{SM}}_{h\rightarrow \xi\xi}(m_0^2),
\end{eqnarray}
where $v=246$ GeV is the SM Higgs doublet vacuum expectation value,
$m_0$ denotes the mass of $0^{++}$, the expression for $y^2/M^2$
depends on the model and will be given in Eq.~\eqref{eqn:ysqovermsq},
and $\Gamma^{\text{SM}}_{h\rightarrow \xi\xi}(m_0^2)$ is the partial
decay width of a SM-like Higgs boson with mass $m_0$ into a pair of
$\xi$'s calculated with HDECAY 6.52
\cite{Djouadi:1997yw,Djouadi:2018xqq}
\footnote{The current limit on the decay width of a scalar boson of mass between 10--60 GeV, which mixes with the SM Higgs boson, comes from the LEP search of the SM-like Higgs boson through $h \to b \bar b$ and $h\to \tau\tau$ decay modes \cite{Barate:2003sz}. The upper bounds on the decay width of $0^{++}$ are approximately $1\%$ of a SM-like Higgs boson's decay width of the same mass $m_0$. Thus, the corresponding lower limits on $c\, \tau_{0^{++}}$ are many orders of magnitude smaller than those presented in Fig.~\ref{fig:lifetimeglueballs}.}.
$0^{++} \rightarrow \xi\xi$ here includes all the decay modes of a SM-like Higgs boson of mass $m_0$ into a pair of SM particles that are allowed kinematically such as $b\bar b$ and $\tau^+ \tau^-$. For the considered mass range of $0^{++}$ i.e. $10-60$ GeV, the dominant decay mode is a pair of $b-$jets. $\alpha_s^B$ and
$\mathbf{F^S_{0^{++}}}$ are respectively the mirror strong coupling
and the mirror glueball annihilation matrix element with $4\pi
\alpha_s^B \mathbf{F^S_{0^{++}}}\approx 2.3 \, m_0^3$
\cite{Curtin:2015fna}.
The expression for $y^2/M^2$ is a compact notation for parameters in
various models of neutral naturalness \cite{Alipour-Fard:2018lsf}:
\begin{eqnarray}\label{eqn:ysqovermsq}
\frac{y^2}{M^2}\approx 
\begin{cases}
\frac{1}{4v^2}\frac{m^2_t}{m^2_{\tilde{t}}},\text{   Folded SUSY,} \;\\
-\frac{1}{2v^2}\frac{m_t^2}{m_T^2},\text{  Fraternal Twin Higgs and Quirky Little Higgs,}\; \\
\frac{1}{2v^2}\frac{v}{v_H}\sin{\theta},\text{   Hyperbolic Higgs,}\;
\end{cases}
\end{eqnarray}
where $m_t$ is the SM top-quark mass, $m_{\tilde{t}}$ is the stop mass,
$m_T$ is the top partner mass, and $v_H$ is the hyperbolic scale and
$\tan{\theta}\approx v/v_H$ encodes the tree-level mixing effects
induced by the mixing between $CP-$even neutral scalars.

As in the discussion on the Higgs-portal model, we present a plot for the proper lifetime as a function of the model parameters. In Fig.~\ref{fig:lifetimeglueballs}, we show isocurves of Log$_{10}(c\tau_{0^{++}})$ in meter in the plane $m_{\tilde{t}/T}$ vs. $m_0$. The plot is in the same format as the corresponding plot in Ref.~\cite{Curtin:2015fna}, but the factor-2 correction to $y^2/M^2$ is taken here into account and leads to changes in the plot. For smaller $m_0$ and larger $m_{\tilde{t}/T}$, the lightest mirror glueballs have a longer proper lifetime.

As for the production of the lightest mirror glueballs from the SM Higgs
decay, we calculate the relevant branching ratio as
\begin{eqnarray}\label{eqn:h200}
  \text{Br}(h\rightarrow 0^{++}0^{++}) \approx
  \text{Br}(h\rightarrow gg)_{\text{SM}}\cdot
  \Big(\frac{\alpha^B_s(m_h)}{\alpha_s^A(m_h)} \, 2 \, v^2
  \Big[  \frac{y^2}{M^2}   \Big]\Big)^2  \cdot
  \sqrt{1-\frac{4m_0^2}{m_h^2}}\cdot \kappa(m_0),
\end{eqnarray}
where $\text{Br}(h\rightarrow gg)_{\text{SM}}\approx 8.6 \, \%$ is the decay branching
ratio of the SM Higgs boson $h$ into a pair of gluons, and
$\frac{\alpha^B_s(m_h)}{\alpha_s^A(m_h)} \sim \mathcal{O}(1)$ is the
ratio of the couplings of the hidden and SM QCD sectors. \footnote{Note that in Eq.~\eqref{eqn:ysqovermsq} and Eq.~\eqref{eqn:h200} there is a factor of 2 difference from the corresponding formulas given in Ref.~\cite{Curtin:2015fna}. This is because in Eq. (3.3) of Ref.~\cite{Curtin:2015fna} an overall factor 1/2 is missing compared to Eq.~(2.2) of Ref.~\cite{Juknevich:2009gg}, the notation of which was adopted in Ref.~\cite{Curtin:2015fna}.}
For a conservative estimate of $\alpha^B_s(m_h)/\alpha_s^A(m_h)$, we extract the
information from the lower green curve of Fig. 5 of
Ref.~\cite{Curtin:2015fna} for $(\alpha^B_s(m_h)/\alpha_s^A(m_h))^2$,
where the one-loop RGE extrapolation from $m_0$ was used assuming
that mirror glueballs are the only mirror states below $m_h$.
$\kappa$ is a parameter taking into account the effect of the glueball hadronization and non-perturbative mixing effects between the excited glueball states $0^{(++*)}$ and the SM Higgs boson.
Following Ref.\cite{Curtin:2015fna},
the maximal value $\kappa_{\text{max}}=1$ is the most 
optimistic signal estimate while the minimal value $\kappa_{\text{min}}$
is the most pessimistic case, which can be estimated under
democratic Higgs-decay principle as follows:
\begin{eqnarray}
  \kappa_{\text{min}}(m_0)=\frac{\sqrt{1-\frac{4m_0^2}{m_h^2}}}
        {\sum\limits_i \sqrt{1-\frac{4m_i^2}{m_h^2}}},
\end{eqnarray}
where $i$ runs over the active states among the 12 stable glueball
states, since for relatively large values of $m_0$, some heavier
mirror-glueball states are forbidden to be produced from the Higgs decay.
This is because in the mirror-glueball spectrum only the ratios
$m_i/m_0$ are known \cite{Morningstar:1999rf,Juknevich:2009gg}.

As in the Higgs-portal model, we take into account the constraint on the parameter space from the invisible Higgs decay width measurement. We find that for $m_{\tilde{t}/T}$ as small as $100-200$ GeV, $\text{Br}(h\rightarrow 0^{++}0^{++})$ is at the largest at $1\%$ level, far below the present experimental upper bound $19\%$. Heavier stop masses would give a branching ratio orders of magnitude below the current bound. Therefore, we conclude that this constraint is irrelevant for the parameter region of this model considered in this work.

\section{Detector setups, simulation \& calculation}\label{sec:simulation}

By simulating 100k events for every parameter point with the Monte-Carlo simulation tool \texttt{Pythia 8.235} we
obtain the total number of signal events $N_{\text{s.e.}}$ by
estimating the number of reconstructed displaced vertices in the IT,
HCAL, or MS.
We switch on ``HiggsSM:all'' of the ``HiggsProcess''
module in \texttt{Pythia 8} to turn on all three Higgs production
processes at an $e^- e^+$ collider, and set the SM Higgs boson to
decay solely into a pair of new scalars, which decay further to a
specified final state depending on which model we are studying.
We place different requirements on the secondary vertices depending on
the detector component and the model, in order to perform the
estimation. Since in each signal event there are two displaced
vertices, we require at least one displaced vertex to be reconstructed
in the IT in order to constitute a signal event, while for HCAL/MS we
require both two vertices reconstructed inside the corresponding
component.  We may express $N_{\text{s.e.}}$ for the IT, HCAL and MS
with the following formulas, respectively,
\begin{eqnarray}
  N_{\text{s.e.}}^{\text{IT}}&=&\mathcal{L}_{h}\cdot \sigma_{h}\cdot
  \text{Br}(h\rightarrow X X) \cdot \left\langle P[s.e.\text{ in IT}]
  \right\rangle\cdot \epsilon^{\text{IT}}, \label{eq:LLPSEIT}\\
  N_{\text{s.e.}}^{\text{HCAL}}&=&\mathcal{L}_{h}\cdot \sigma_{h}\cdot
  \text{Br}(h\rightarrow X X) \cdot \left\langle P[s.e.\text{ in HCAL}]
  \right\rangle, \label{eq:LLPSEHCAL}\\
  N_{\text{s.e.}}^{\text{MS}}&=&\mathcal{L}_{h}\cdot \sigma_{h}\cdot
  \text{Br}(h\rightarrow X X) \cdot \left\langle P[s.e.\text{ in MS}]
  \right\rangle\;. \label{eq:LLPSEMS}
\end{eqnarray}
Here $\mathcal{L}_h$ is the integrated luminosity at the Higgs mode,
$\sigma_{h}$ is the total cross section for the SM Higgs production by
combining the three processes at the $e^- e^+$ colliders ($e^- e^+
\rightarrow HZ$ and two vector-boson-fusion processes: $WW$ and $ZZ$),
$X$ represents either the light scalar boson $h_s$ or the lightest
mirror glueball $0^{++}$, and $\epsilon^{\text{IT} }$ denotes the cut
efficiency for the IT. For both the CEPC and FCC-ee, the number of the
Higgs bosons produced
$N_h = \mathcal{L}_h \cdot \sigma_{h}\simeq 1.14\times 10^6$
for operation at the Higgs mode with $\sqrt{s}=240$
GeV \cite{CEPCStudyGroup:2018rmc,Abada:2019zxq}.  $\left\langle
P[s.e.\text{ in IT}] \right\rangle$ denotes the average probability for
at least one of the LLPs decaying inside the IT. $\left\langle
P[s.e.\text{ in HCAL}] \right\rangle$ and $\left\langle P[s.e.\text{
    in MS}] \right\rangle$ are similar notations employed for the HCAL
and MS, respectively. These average decay probabilities may be
calculated as follows:
\begin{eqnarray}
  \left\langle P[s.e.\text{ in IT}] \right\rangle&=&\frac{1}{N^{\text{MC}}}
  \sum_{i=1}^{N^\text{MC}}\bigg( P[X^1_{i}\text{ in IT}]  +
  P[X^2_{i}\text{ in IT}] -  P[X^1_{i}\text{ in IT}]  \cdot
  P[X^2_{i}\text{ in IT}]  \bigg), \label{eq:averagedecayprobIT}\nonumber\\
  \left\langle P[s.e.\text{ in HCAL}] \right\rangle&=&
  \frac{1}{N^{\text{MC}}}\sum_{i=1}^{N^\text{MC}}\bigg(  P[X^1_{i}\text{ in HCAL}]
  \cdot  P[X^2_{i}\text{ in HCAL}]  \bigg), \label{eq:averagedecayprobHCAL}\nonumber\\
  \left\langle P[s.e.\text{ in MS}] \right\rangle&=&\frac{1}{N^{\text{MC}}}
  \sum_{i=1}^{N^\text{MC}}\bigg(  P[X^1_{i}\text{ in MS}]  \cdot
  P[X^2_{i}\text{ in MS}]  \bigg), \label{eq:averagedecayprobMS}
\end{eqnarray}
where $N^{\text{MC}}$ is the total number of MC-simulated events, and
$P[X_i^{1/2}\text{ in IT/HCAL/MS}]$ is the individual decay probability of the first/second of the two LLPs in the $i$-th simulated signal event inside the respective fiducial component. Before we show how to calculate the latter, we first introduce the geometries of the IT, HCAL, and MS.

\begin{figure}
	\includegraphics[scale=0.5]{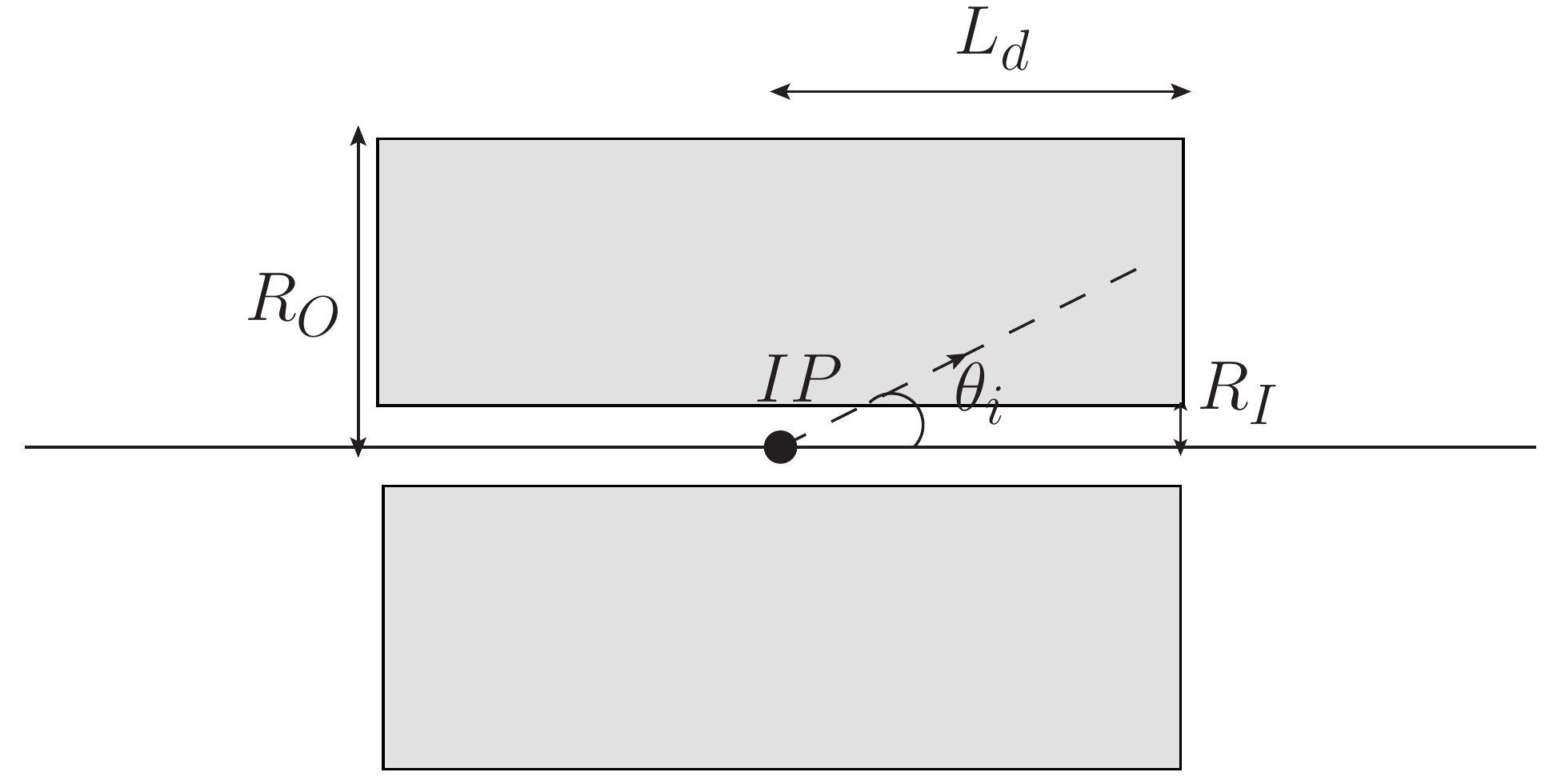}
	\includegraphics[scale=0.5]{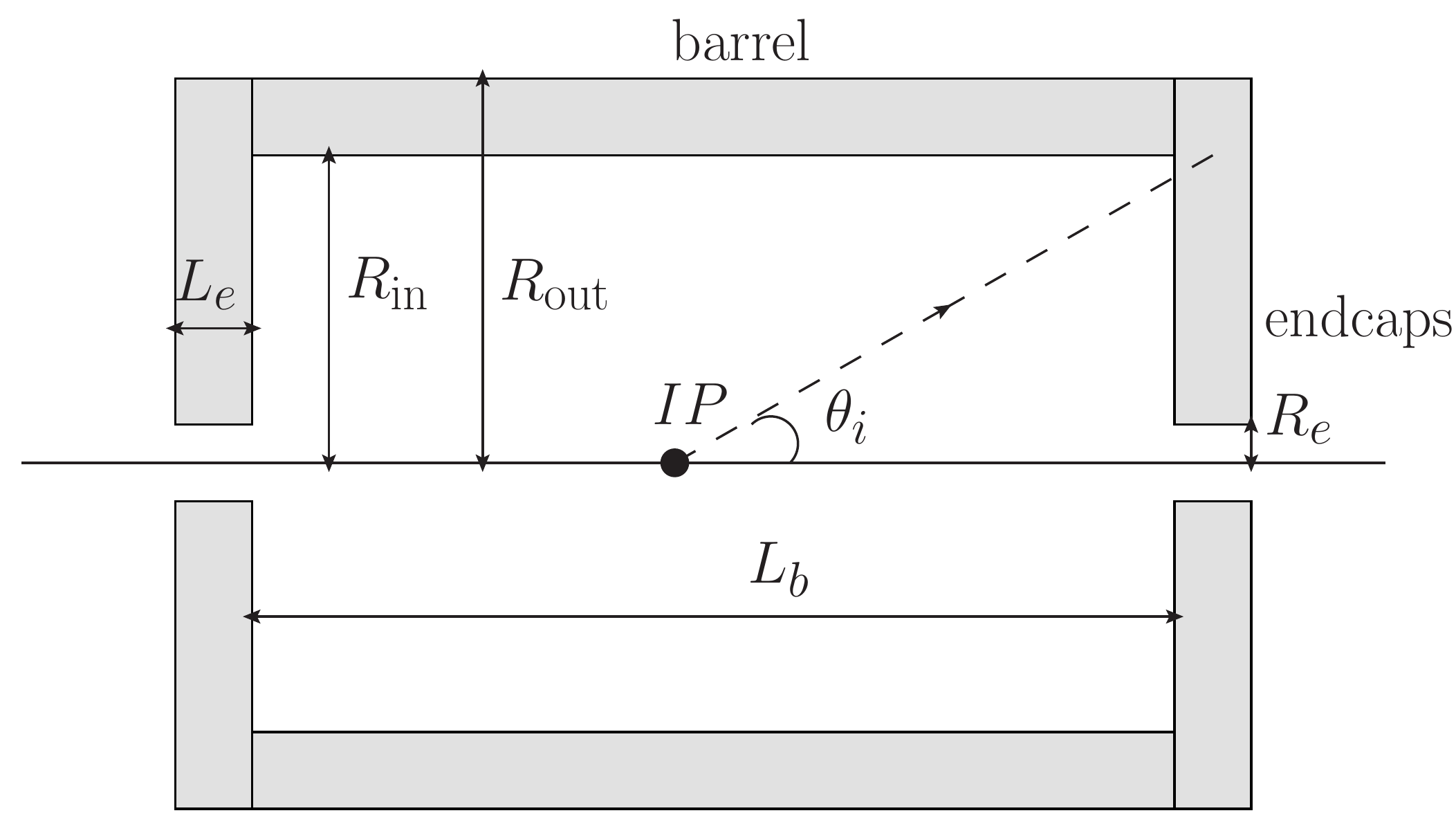}
	\caption{Profile sketch of the detector components of the CEPC and FCC-ee. The upper plot is for the the inner tracker and the lower one is for the HCAL and muon chamber. $\theta_i$ labels the polar angle of an example LLP with its traveling path depicted by the accompanying dashed line with an arrow.}
	\label{fig:detectorsketch}
\end{figure}

A profile sketch of the CEPC and FCC-ee detectors is shown in Fig.~\ref{fig:detectorsketch}.
The setups in the CEPC baseline detector and FCC-ee IDEA design are similar
\cite{CEPCStudyGroup:2018ghi,Benedikt:2018qee}.
Summaries of the geometrical parameters used for the IT, HCAL, and MS in
this work extracted from Refs.~\cite{CEPCStudyGroup:2018ghi,Benedikt:2018qee} are listed in Table~\ref{tab:innerparameters}, Table~\ref{tab:hcalparameters}, and Table~\ref{tab:muonchamberparameters}, respectively.
We now describe each fiducial volume for the calculation of $P[X_i^{1/2}\text{ in IT/HCAL/MS}]$.
For leptonic final states ($h_s\rightarrow \mu^- \mu^+$) we consider
both the IT and MS for reconstructing the displaced vertices, while for
the hadronic final states ($h_s \rightarrow \pi^- \pi^+$ and
$0^{++}\rightarrow b\bar{b}$) we further include the HCAL as a part of the
fiducial volume.
For the IT we consider the vertex detector and the silicon inner tracker.
The HCAL and muon spectrometer consist of a barrel and two endcaps each.

For signal events in the IT, we require at least
one displaced vertex in each event. Thus, the expression
for $P[X_i\text{ in IT}]$ is given by
%in Eq.~\eqref{eqn:individualdecprobinnerdetector}.
\begin{eqnarray}
\label{eqn:individualdecprobinnerdetector}
P[X_i\text{ in IT}] &=& e^{-L_i/\lambda_i^t} \cdot (1-e^{-L'_i/\lambda^t_i}),\\
L_i &\equiv & 
\begin{cases}
R_I, \text{ if } |L_d\,\tan{\theta_i}|\leq R_I\\
d_{\text{res}}, \text{ else,}
\end{cases}\nonumber\\
%L_i &\equiv & \text{min}(L_d,|R_I/\tan{\theta_i}|),\nonumber\\
L'_i &\equiv & \text{min}(\text{max}(R_I, |L_d\,\tan{\theta_i}|),R_O)-L_i, \nonumber\\
\lambda_i^t&=& \beta_i^t \, \gamma_i \, \tau_{X}, \nonumber 
\end{eqnarray}
where $\beta_i^t$ is the speed of the $X_i$ in the transverse direction
{with $\gamma_i$ its boost, and $\tau_{X}$ is the lifetime of $X$.
  $R_I$ ($R_O$) is the inner (outer) radius of
the inner detector, and $L_d$ is its half length. $d_\text{res}=5\, \mu $m
is the inner-tracker spatial resolution for both CEPC and FCC-ee
\cite{CEPCStudyGroup:2018rmc,Abada:2019zxq}. As long as
one of the LLPs travels inside the IT window, and decays before it leaves the IT
(including the case that the secondary vertex is inside the beam
pipe up to $d_{\text{res}}$), we treat the decay vertex
as a displaced vertex that can be reconstructed.
The kinematical variables may be obtained with the following relations:
\begin{eqnarray}
\beta_i^t &=& |p_i^t/E_i|,\\
\gamma_i &=& E_i/m,
\end{eqnarray}
where $p_i^t$, $E_i$ and $m$ are respectively the transverse momentum of $X_i$, its energy, and its mass.

Note that for the HCAL and MS we require both displaced vertices to
be reconstructed, in order to render the event as a signal. The formulas of $P[X_i^{1/2}\text{in HCAL/MS}]$ for
the CEPC and FCC-ee are the same, though the geometrical parameters
are slightly different.  
The expression for $P[X_i\text{ in HCAL/MS}]$ is given by
%Eq.~\eqref{eqn:individualdecprobmuonchamber}.
%
\begin{eqnarray}\label{eqn:individualdecprobmuonchamber}
  P[X_i\text{ in HCAL/MS}] &=& e^{-R_e/\lambda_i^z} \cdot
  (1-e^{-L^\alpha_i/\lambda^t_i}) - e^{-R_e/\lambda_i^z} \cdot
  (1-e^{-L^\beta_i/\lambda^t_i})  ,\\
    L^\alpha_i &\equiv & \text{min}(\text{max}(R_e,
    |(\frac{L_b}{2}+L_e)\,\tan{\theta_i}|),R_{\text{out}})-R_e, \nonumber\\
    L^\beta_i &\equiv & \text{min}(\text{max}(R_e,
    |\frac{L_b}{2}\,\tan{\theta_i}|),R_{\text{in}})-R_e, \nonumber
\end{eqnarray}
where $R_{\textrm{in}}$ ($R_{\textrm{out}}$) is the inner (outer) radius of the
barrel, $L_b$ its full length, and $L_e$ ($R_e$) is the width (inner radius)
of the two endcaps.

\begin{table}
	\begin{center}
		\begin{tabular}{c|c|c|c|c}
			\hline
			\hline
			Detector    & $R_I$ [mm] & $R_O$ [m] & $L_d$ [m] & $V$ [m$^3$]\\
			\hline
			CEPC        & 16       & 1.8    & 2.35     & 47.8  \\
			\hline
			FCC-ee IDEA & 17       &   2.0   &  2.0  &  50.3\\
			\hline
			\hline                              
		\end{tabular}
		\caption{Summary of parameters of the IT of the CEPC and FCC-ee IDEA. The parameters of the CEPC baseline detector are extracted from Refs.~\cite{CEPCStudyGroup:2018ghi} while the geometries of the IDEA detectors of the FCC-ee are reproduced from Ref.~\cite{Benedikt:2018qee}. $V$ represents the volume and similarly in Table~\ref{tab:hcalparameters} and Table~\ref{tab:muonchamberparameters}. 
		}
		\label{tab:innerparameters}
	\end{center}
\end{table}

\begin{table}
	\begin{center}
		\begin{tabular}{c|c|c|c|c|c|c}
			\hline
			\hline
			Detector    & $L_b$ [m] & $L_e$ [m] & $R_e$ [m] & $R_{\mathrm{in}}$ [m] & $R_{\mathrm{out}}$ [m] & $V$ [m$^3$]\\
			\hline
			CEPC        & 5.3       &   1.493   &  0.50 & 2.058 & 3.38   &  224.5\\
			\hline
			FCC-ee IDEA   & 6       & 2.5    & 0.35 & 2.5 & 4.5   & 580.1 \\
			\hline
			\hline                              
		\end{tabular}
		\caption{Summary of parameters of the HCAL of the CEPC and FCC-ee IDEA.
		}
		\label{tab:hcalparameters}
	\end{center}
\end{table}

\begin{table}
	\begin{center}
		\begin{tabular}{c|c|c|c|c|c|c}
			\hline
			\hline
			Detector    & $L_b$ [m] & $L_e$ [m] & $R_e$ [m] & $R_{\mathrm{in}}$ [m] & $R_{\mathrm{out}}$ [m]& $V$ [m$^3$]\\
			\hline
			CEPC        & 8.28       &   1.72   &  0.50 & 4.40 & 6.08 & 854.8   \\
			\hline
			FCC-ee IDEA   & 11       & 1    & 0.35 & 4.5 & 5.5  & 534.9 \\
			\hline
			\hline                              
		\end{tabular}
		\caption{Summary of parameters of the MS of the CEPC and FCC-ee IDEA.
		}
		\label{tab:muonchamberparameters}
	\end{center}
\end{table}

\section{Results}\label{sec:results}

In this section, we present the numerical results of the sensitivity
reach of the CEPC and FCC-ee for the two models considered in this
study. 

\subsection{The light sub-GeV scalar boson case}

\begin{figure}
	\includegraphics[width=0.4\textwidth]{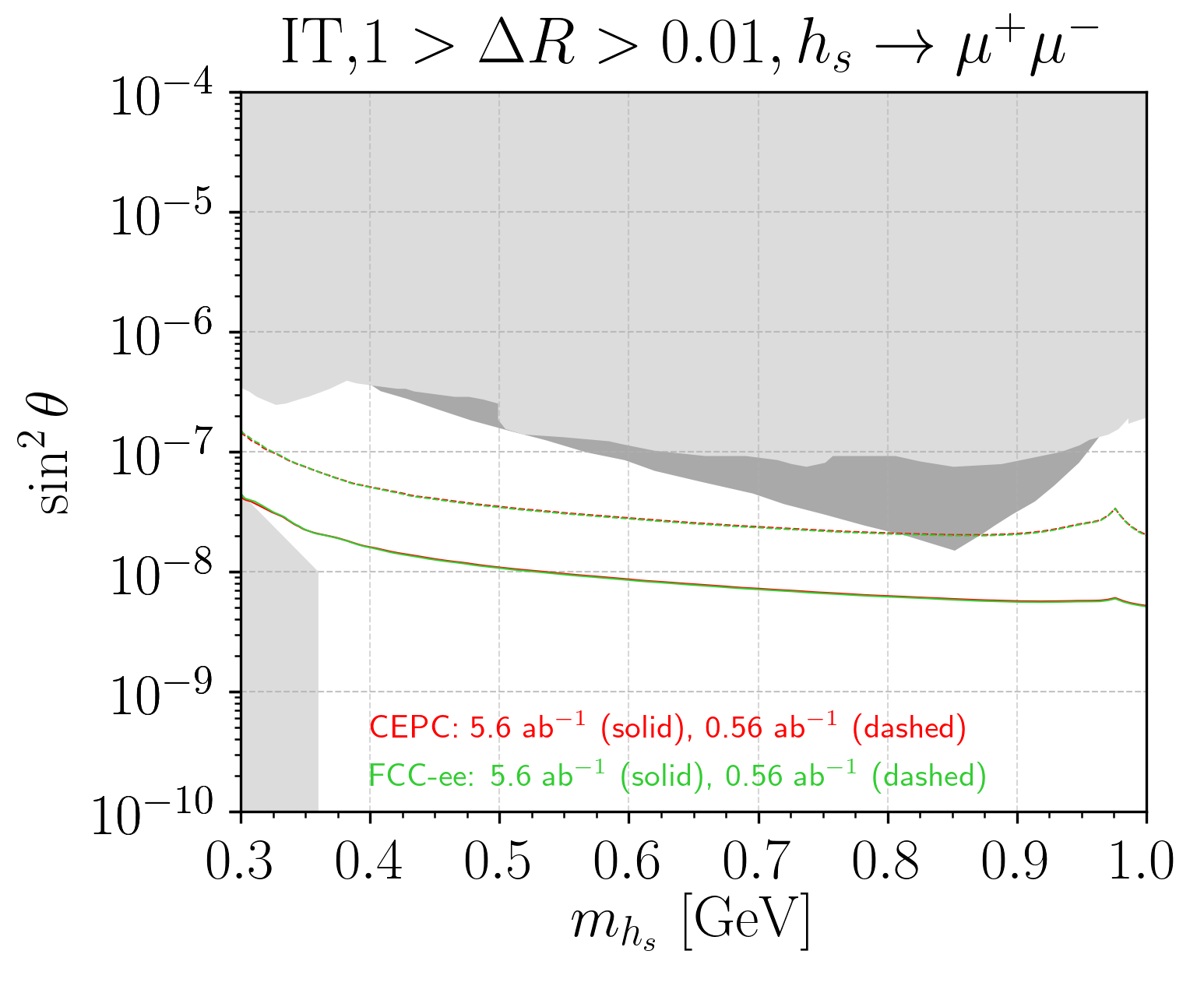}
	\includegraphics[width=0.4\textwidth]{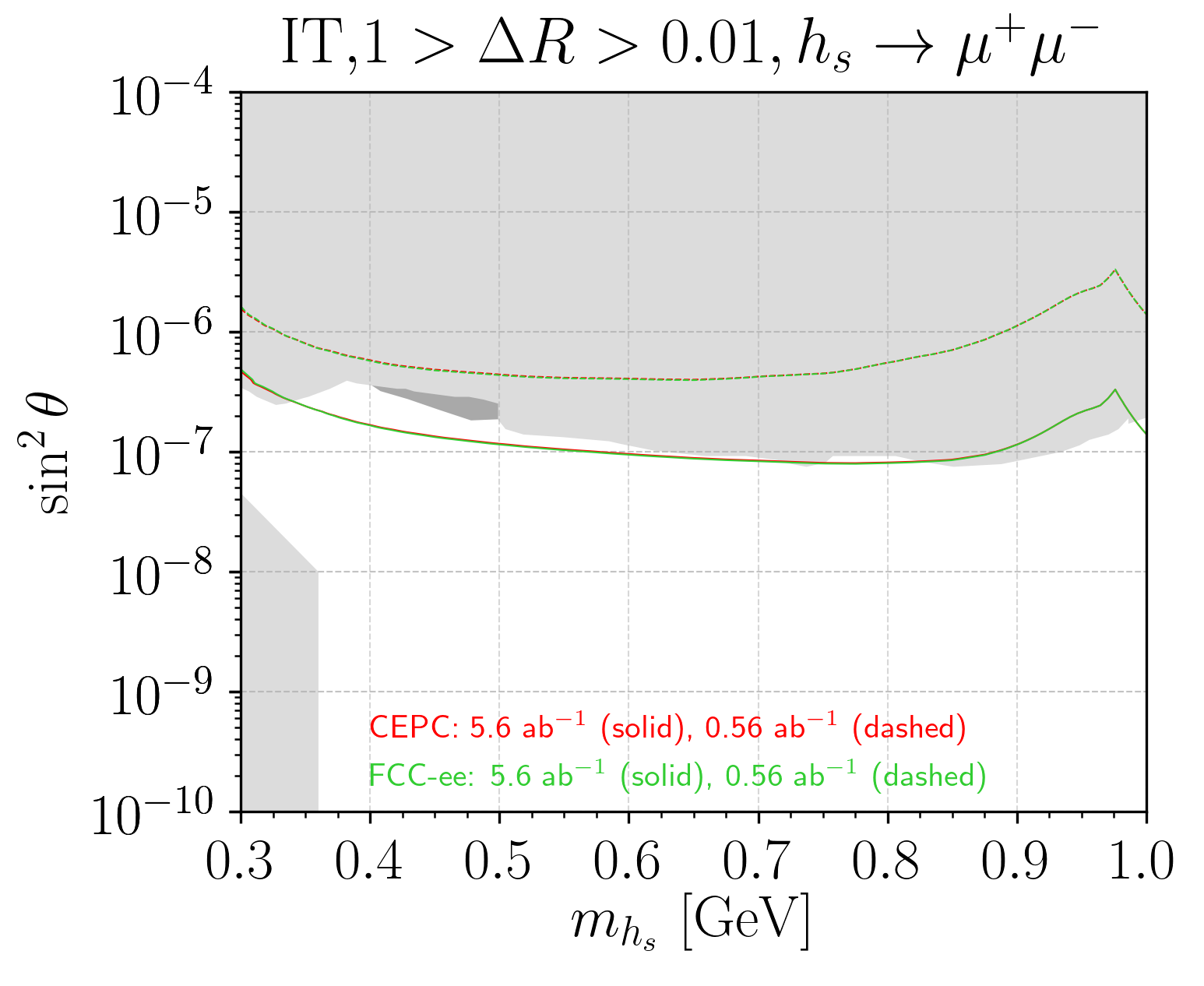}
	\includegraphics[width=0.4\textwidth]{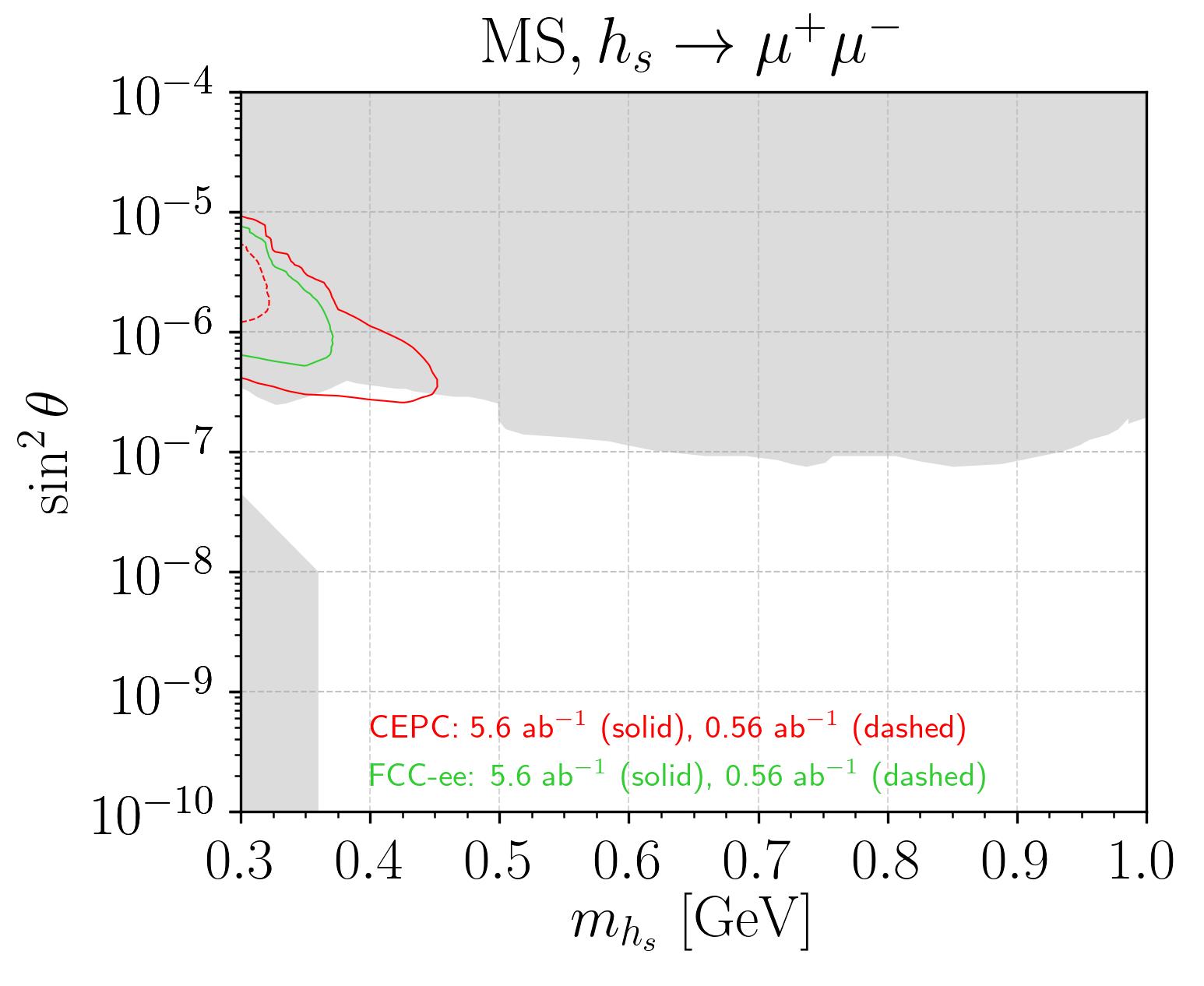}
	\includegraphics[width=0.4\textwidth]{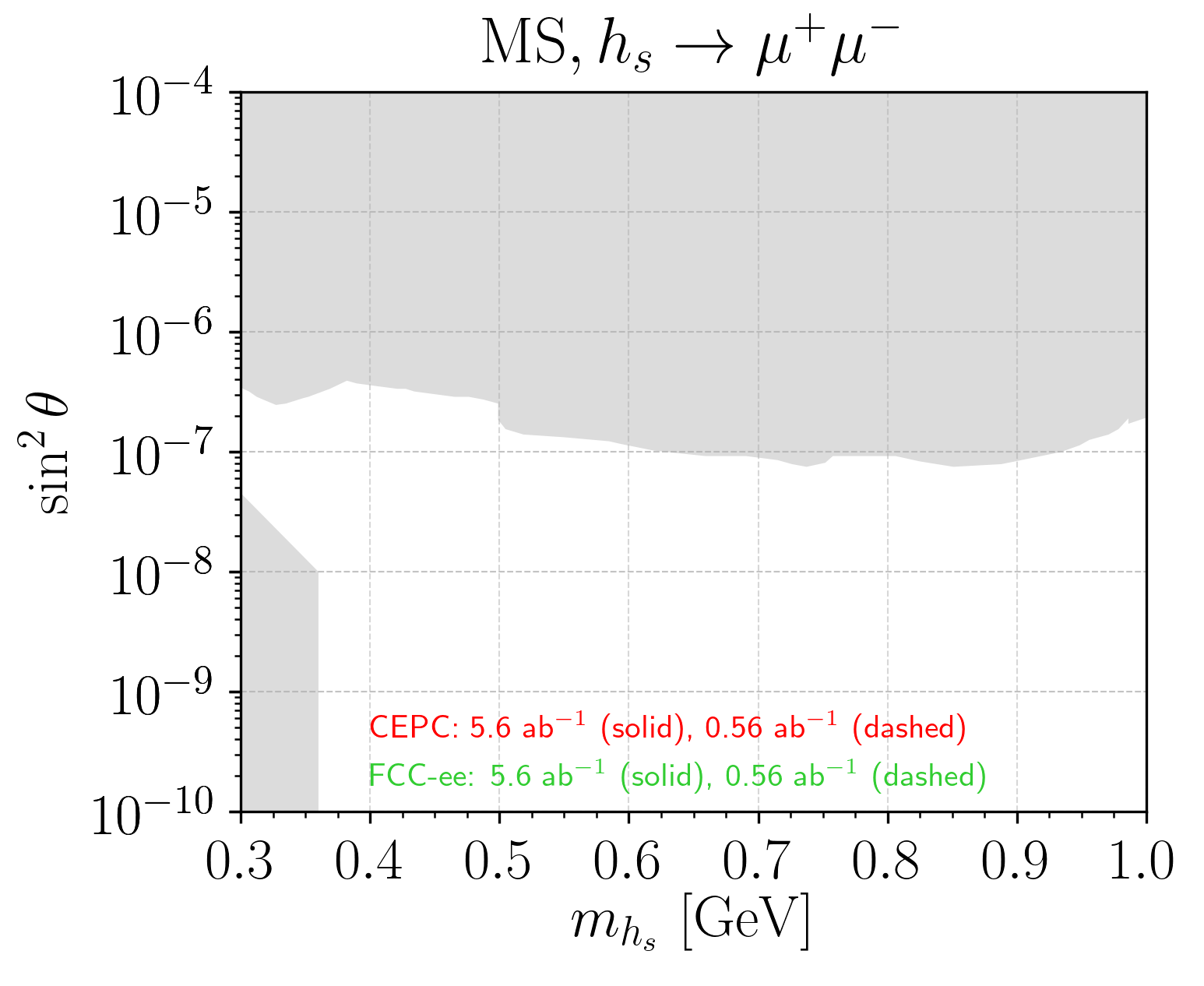}
	\caption{Sensitivity reaches at the CEPC and FCC-ee for
          $h_s\rightarrow \mu^+ \mu^-$.
          The left panels correspond to  $\langle \chi \rangle = 10$ GeV
          while the right ones to $\langle \chi \rangle = 100$ GeV. The light gray area is experimentally excluded while the dark gray part shown in the upper row can be probed at the LHC with 300 fb$^{-1}$ integrated luminosity.}
	\label{fig:sensitivity1}
\end{figure}

\begin{figure}
	\includegraphics[width=0.4\textwidth]{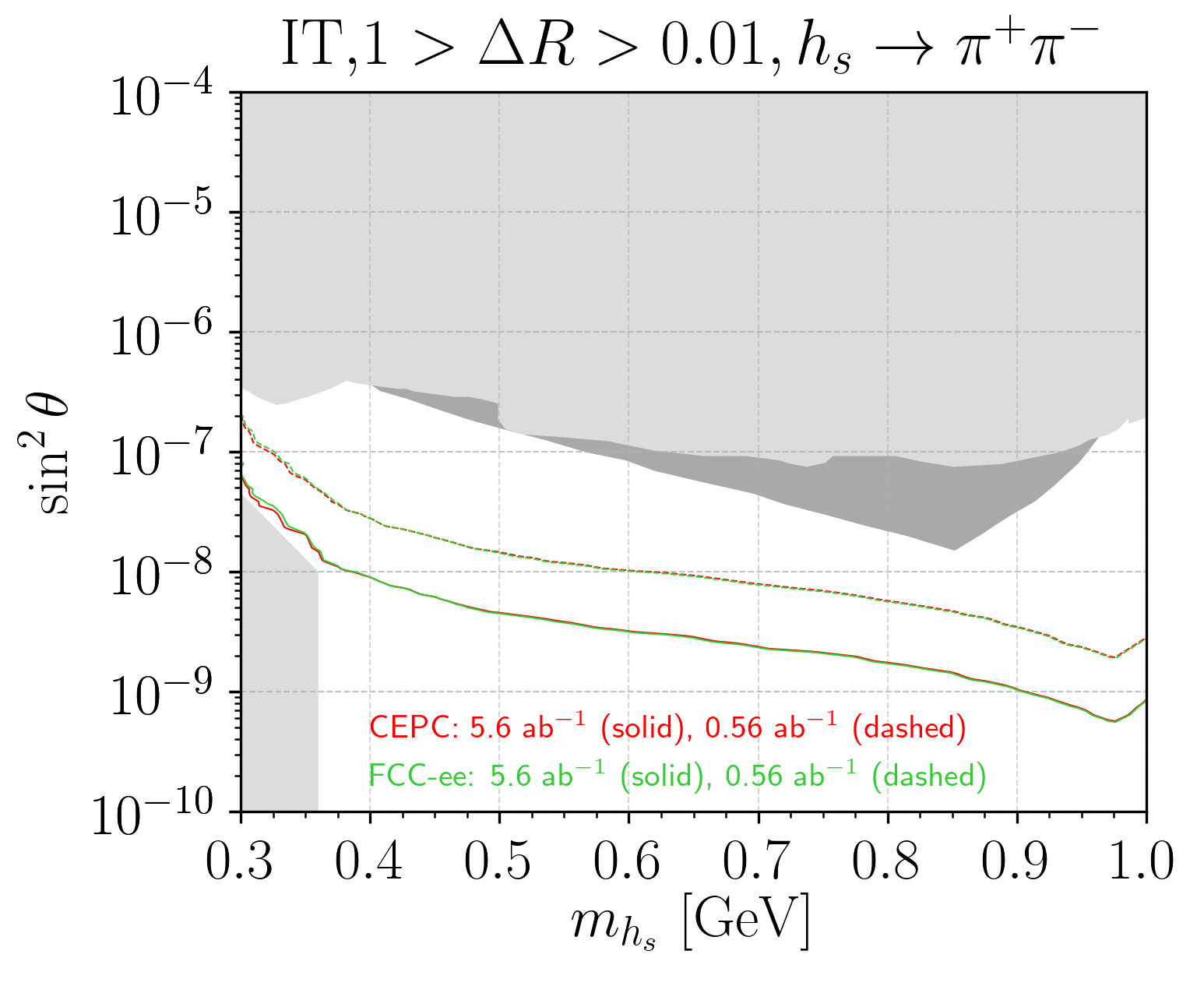}
	\includegraphics[width=0.4\textwidth]{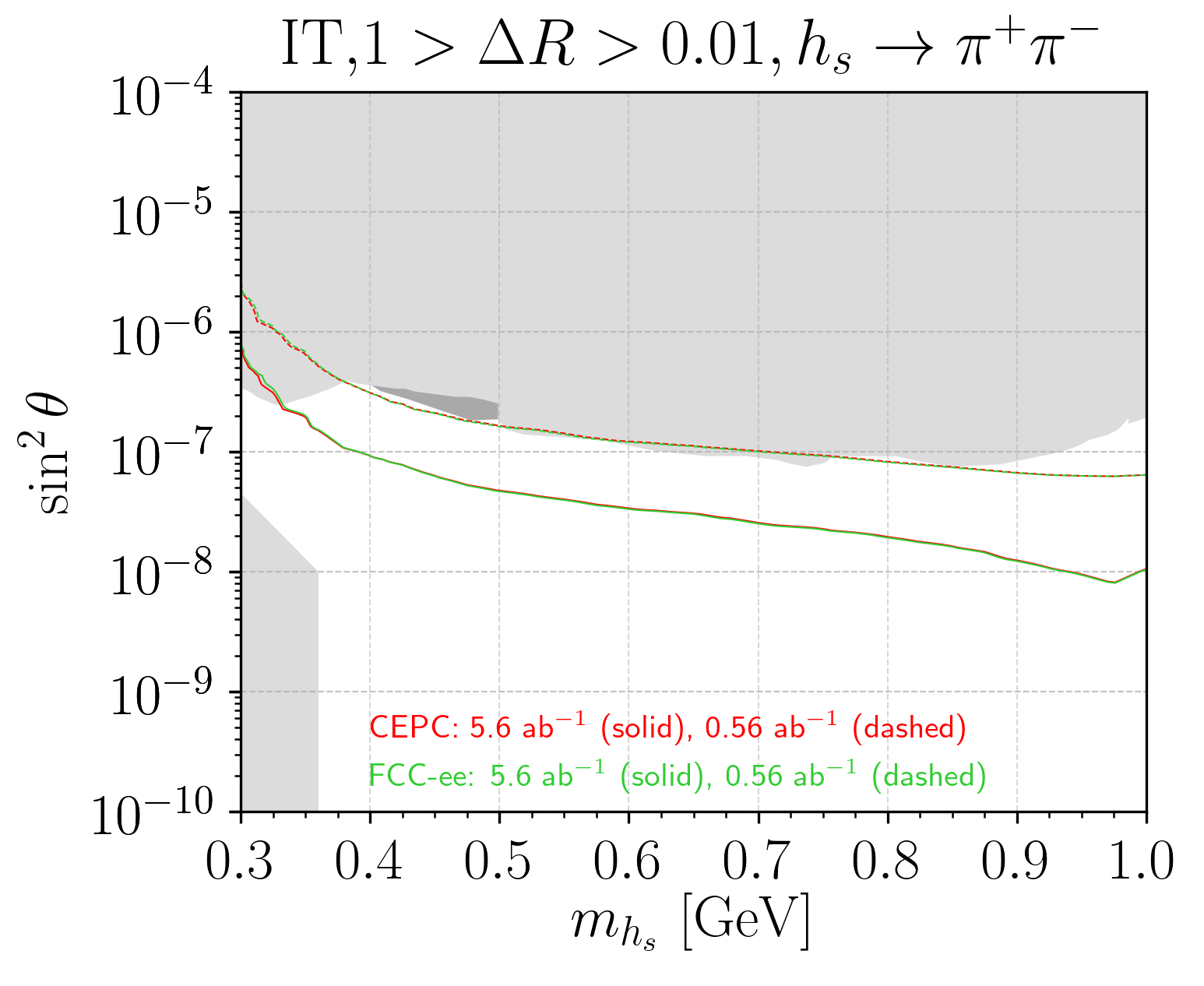}
	\includegraphics[width=0.4\textwidth]{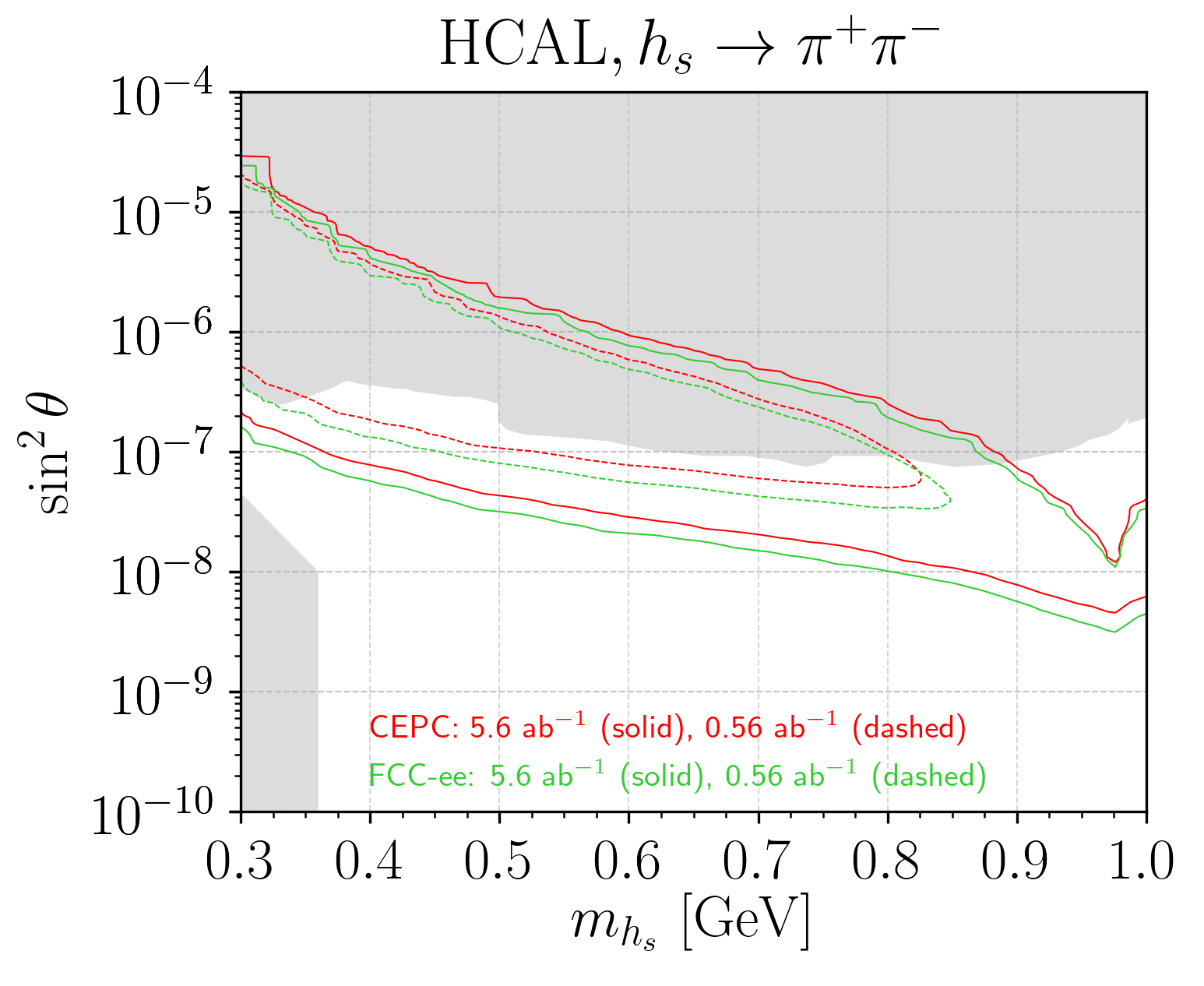}
	\includegraphics[width=0.4\textwidth]{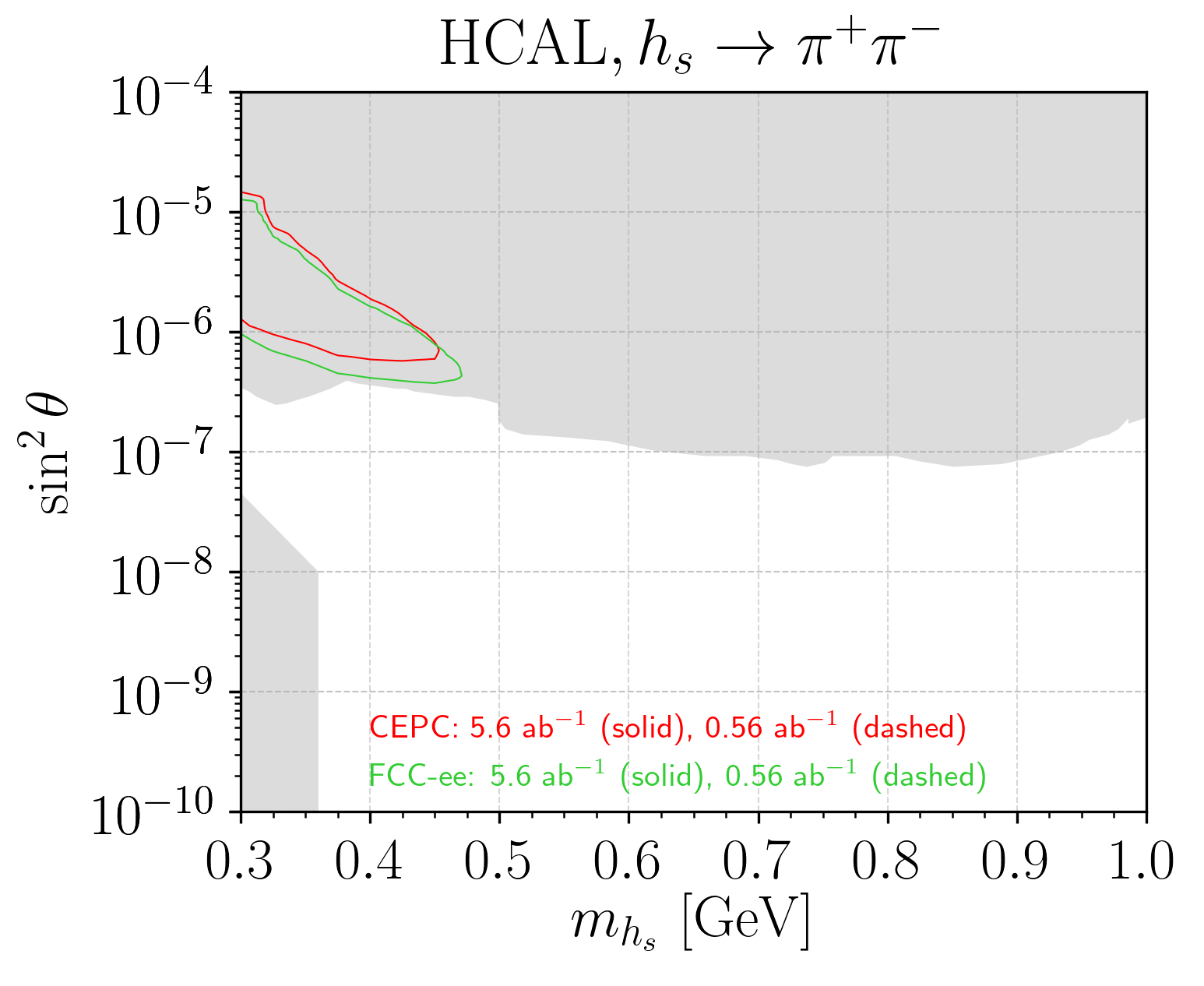}
	\includegraphics[width=0.4\textwidth]{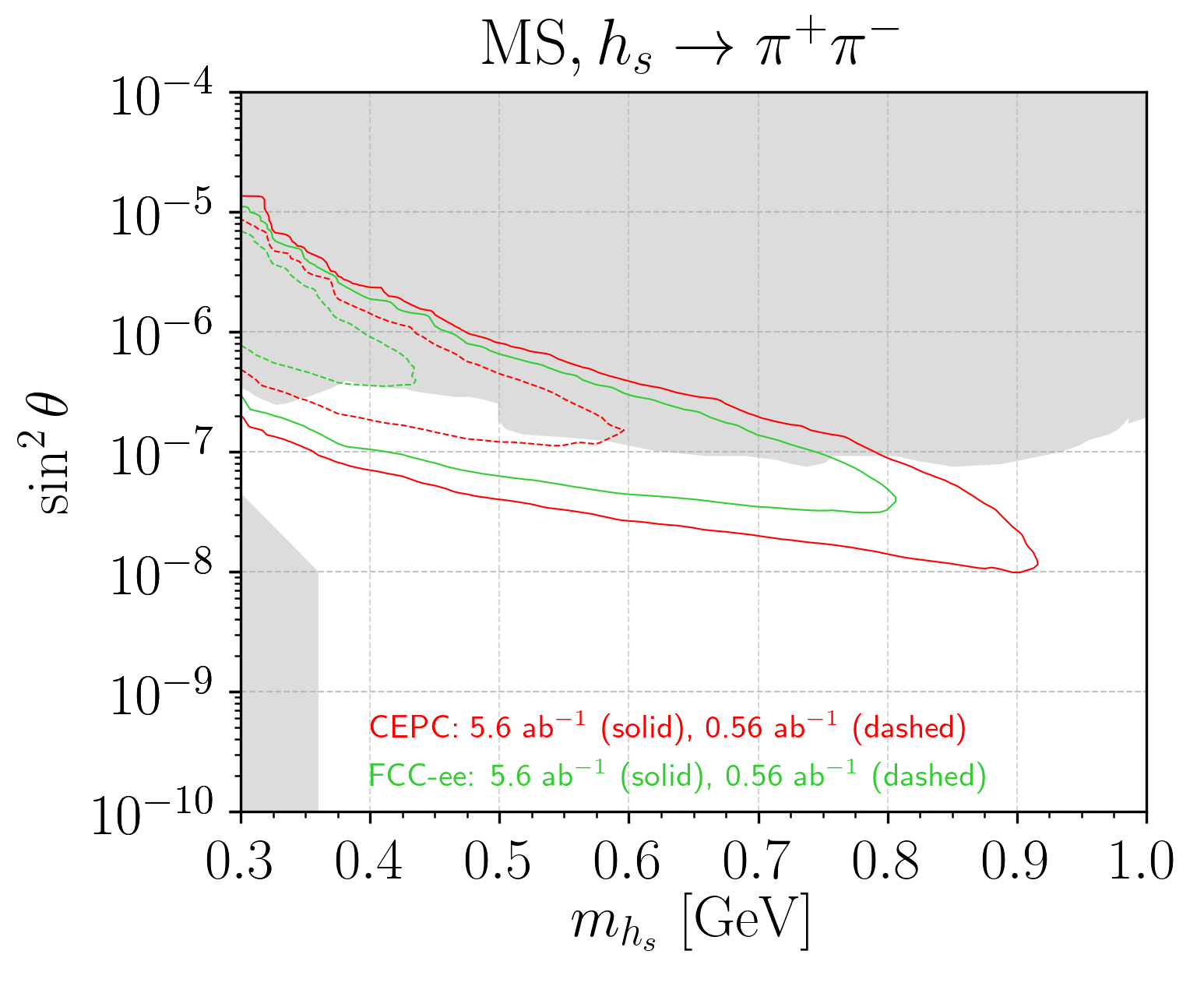}
	\includegraphics[width=0.4\textwidth]{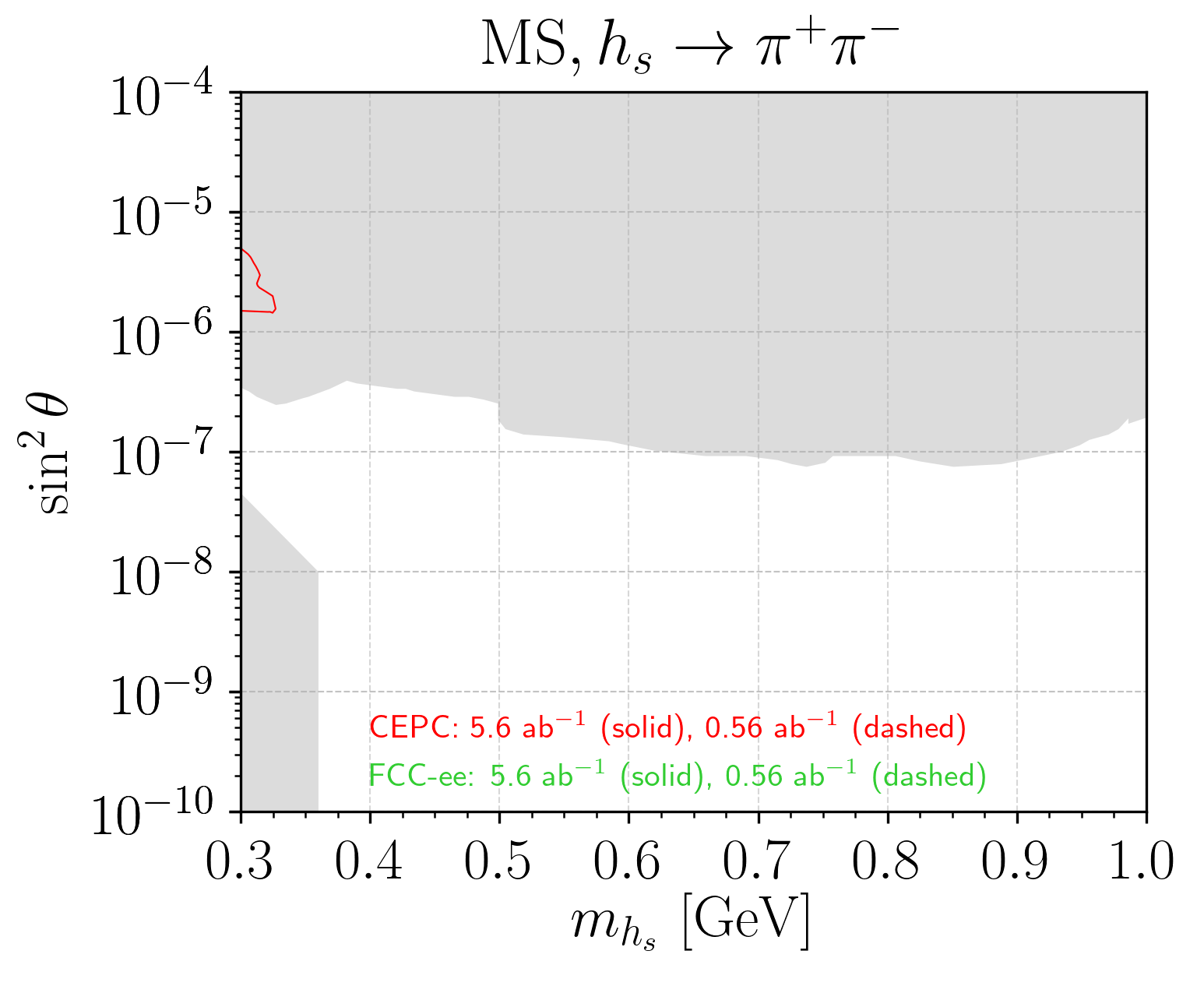}
	\caption{Sensitivity reaches at the CEPC and FCC-ee for
          $h_s \rightarrow \pi^+ \pi^-$.
          The left panels correspond to  $\langle \chi \rangle = 10$ GeV
          while the right ones to $\langle \chi \rangle = 100$ GeV. The light gray area is experimentally excluded while the dark gray part shown in the upper row can be probed at the LHC with 300 fb$^{-1}$ integrated luminosity.}
	\label{fig:sensitivity2}
\end{figure}

We present the sensitivity estimates in the plane ($\sin^2{\theta}$ vs
$m_{h_s}$)
for the light sub-GeV scalar-boson model in Fig.~\ref{fig:sensitivity1}
and Fig.~\ref{fig:sensitivity2} for the channels
$h_s \rightarrow \mu^- \mu^+$ and $h_s \rightarrow \pi^- \pi^+$, respectively.
We consider the number of signal events larger or equal to 3 as
the sensitive region, which corresponds to the exclusion limit of 95\%
C.L. with 0 background event.
The ITs of the CEPC and FCC-ee employ
similar geometries and consequently their results are almost
identical, while the HCAL/MS of the two detectors somewhat differ in their dimensions and hence volumes, leading to slightly different results.
In Figs.~\ref{fig:sensitivity1} and \ref{fig:sensitivity2}, 
the solid lines represent the projected integrated luminosity of 
$\mathcal{L}_h=5.6$ ab$^{-1}$ while the dashed ones are for
$\mathcal{L}_h=0.56$ ab$^{-1}$ only.
We show results for these two integrated luminosities in order to achieve
fair comparisons with the results in Ref.~\cite{Chang:2016lfq},
where the LHC sensitivities for the same model with dimuon displaced
vertices were obtained for an integrated luminosity of 300 fb$^{-1}$
while at the end of high-luminosity LHC (HL-LHC) run 3 ab$^{-1}$ of data
is expected.
We apply a selection cut on the opening angle of the two muons/pions
produced from the $h_s$ decays inside the IT: $1 > \Delta R > 0.01$. This can
effectively eliminate the contributions from background events of
heavier particles while respecting the tracking spatial resolution.
As for the HCAL and MS, we assume 0 background events.
The LHC results from Ref.~\cite{Chang:2016lfq} are reproduced here in
Fig.~\ref{fig:sensitivity1} and Fig.~\ref{fig:sensitivity2} for easy
comparisons. The two benchmark values of $\langle \chi \rangle = 10$, 100 GeV are chosen as we consider light $h_s$ of sub-GeV mass. The light gray area is the experimentally excluded region by fixed-target experiments, LHCb, and B-factories \cite{Clarke:2013aya,Aaij:2015tna,Chang:2016lfq,Aaij:2016qsm}, while the dark gray area is the experimentally allowed and LHC-sensitive region \cite{Chang:2016lfq}. Note that the strongest limits for $m_{h_s}$ between 0.3 and 1.0 GeV were obtained by the LHCb searches \cite{Aaij:2015tna,Aaij:2016qsm} for $B-$meson decays. Therefore we refrain from presenting a comprehensive overview of the previous searches which had weaker limits for the relevant mass range. See the scalar-portal discussion in Ref.~\cite{Beacham:2019nyx} for a summary of the update-to-date experimental constraints for $m_{h_s}\lsim10$ GeV.

For the dimuon signature case, with the smaller $\langle \chi \rangle =10$ GeV, the CEPC and FCC-ee with an integrated luminosity of 0.56 ab$^{-1}$
are slightly weaker at $\sim m_{h_s}=0.85$ GeV, but can have as much
as a two-orders-of-magnitude advantage in the rest of the considered
$m_{h_s}$ range.
As for $\langle \chi \rangle=100$ GeV, the performance of the $e^- e^+$ 
colliders with an integrated luminosity of 0.56 ab$^{-1}$ is
inferior to the LHC for most of the mass range. Furthermore, with the integrated
luminosity of 5.6 ab$^{-1}$ the performance is better than the LHC
with 300 fb$^{-1}$ integrated luminosity
for both small and large $\langle \chi \rangle$ values.
While the sensitivity reach of the IT is
dominant, our MS estimates still show some potential in the parameter
space, at least in the $\langle\chi\rangle=10$ GeV case, though for
$\langle\chi\rangle=100$ GeV the MS has no sensitivity in the
considered mass range as there would be too few $h_s$ produced.

On the other hand, by including the signatures from displaced charged pions
one may significantly enhance the signal sensitivities, as for
$m_{h_s}\lsim 1$ GeV the light scalar boson almost solely decays
into (neutral or charged) pions. We expect that collimated pions can be detected
thanks to the clean environment at $e^- e^+$ colliders.
The results are shown in
Fig.~\ref{fig:sensitivity2}. Roughly for $m_{h_s}\gsim 0.7$ GeV, we
observe that the sensitivities improve by orders of magnitude compared
to those in the dimuon signature case. Now the CEPC/FCC-ee IT may
explore $\sin^2{\theta}$ a few orders of magnitude lower than the sensitive region of the LHC for the whole mass range.
In fact, for $\langle\chi\rangle=10$ GeV even the HCAL with $\mathcal{L}=0.56$
ab$^{-1}$ may have a comparable reach to that of the LHC.
Similar to the dimuon signature, for $\langle\chi\rangle=100$ GeV the
sensitivities at the HCAL and MS are worsened by a large extent because
of the reduced production cross section of $h_s$.

\subsection{The Mirror Glueball case}
\begin{figure}
	\includegraphics[width=0.4\textwidth]{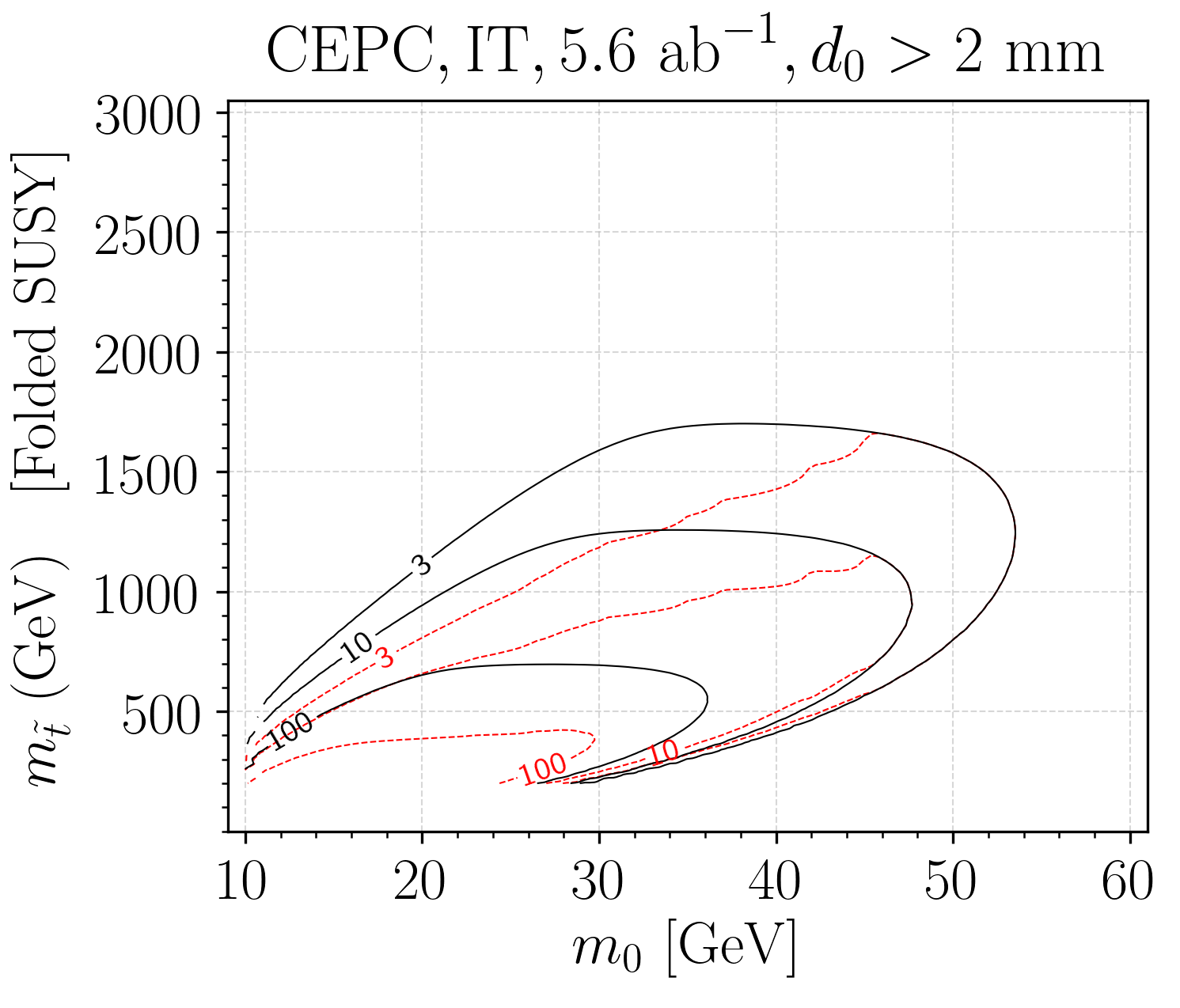}
	\includegraphics[width=0.4\textwidth]{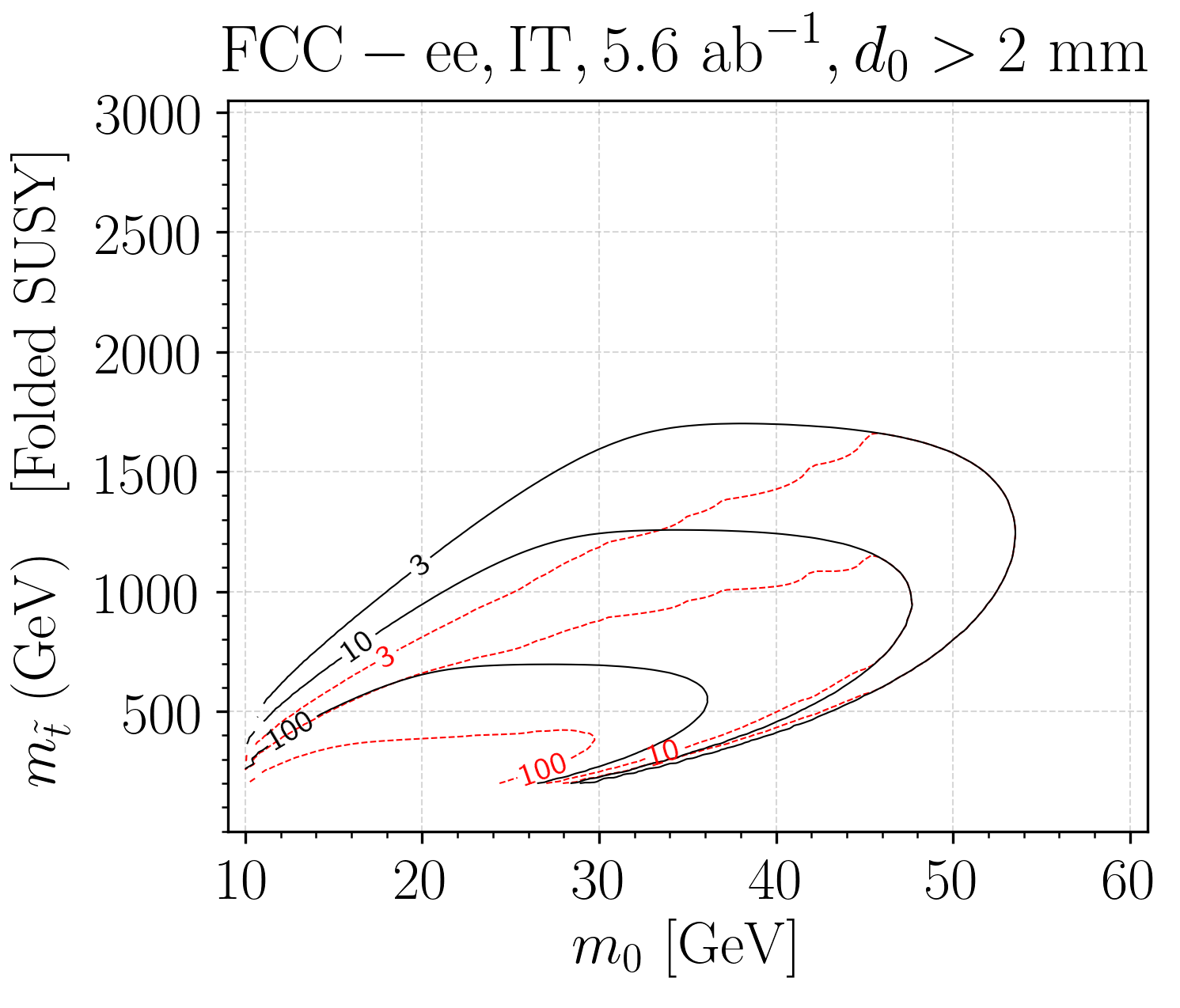}
	\includegraphics[width=0.4\textwidth]{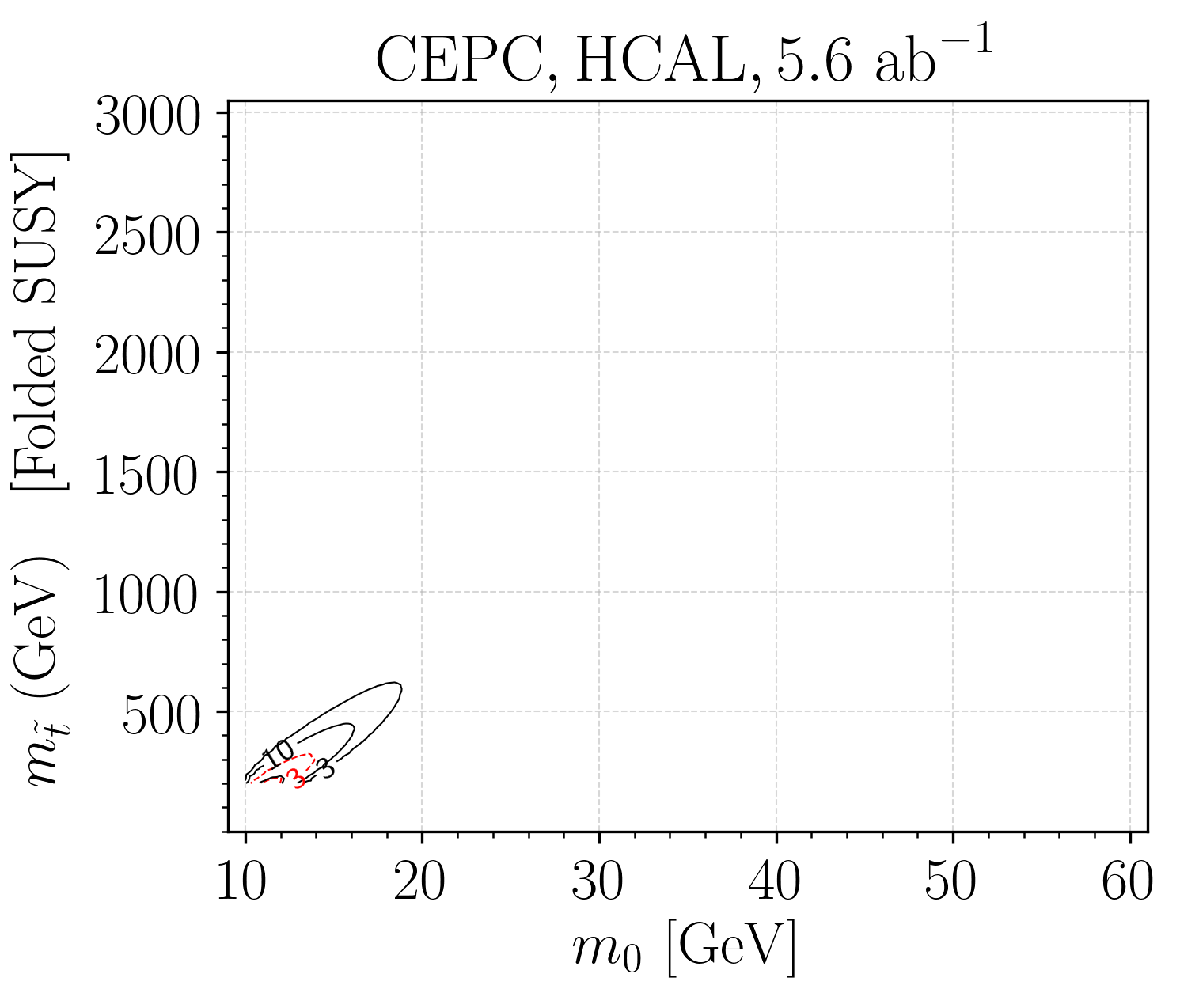}
	\includegraphics[width=0.4\textwidth]{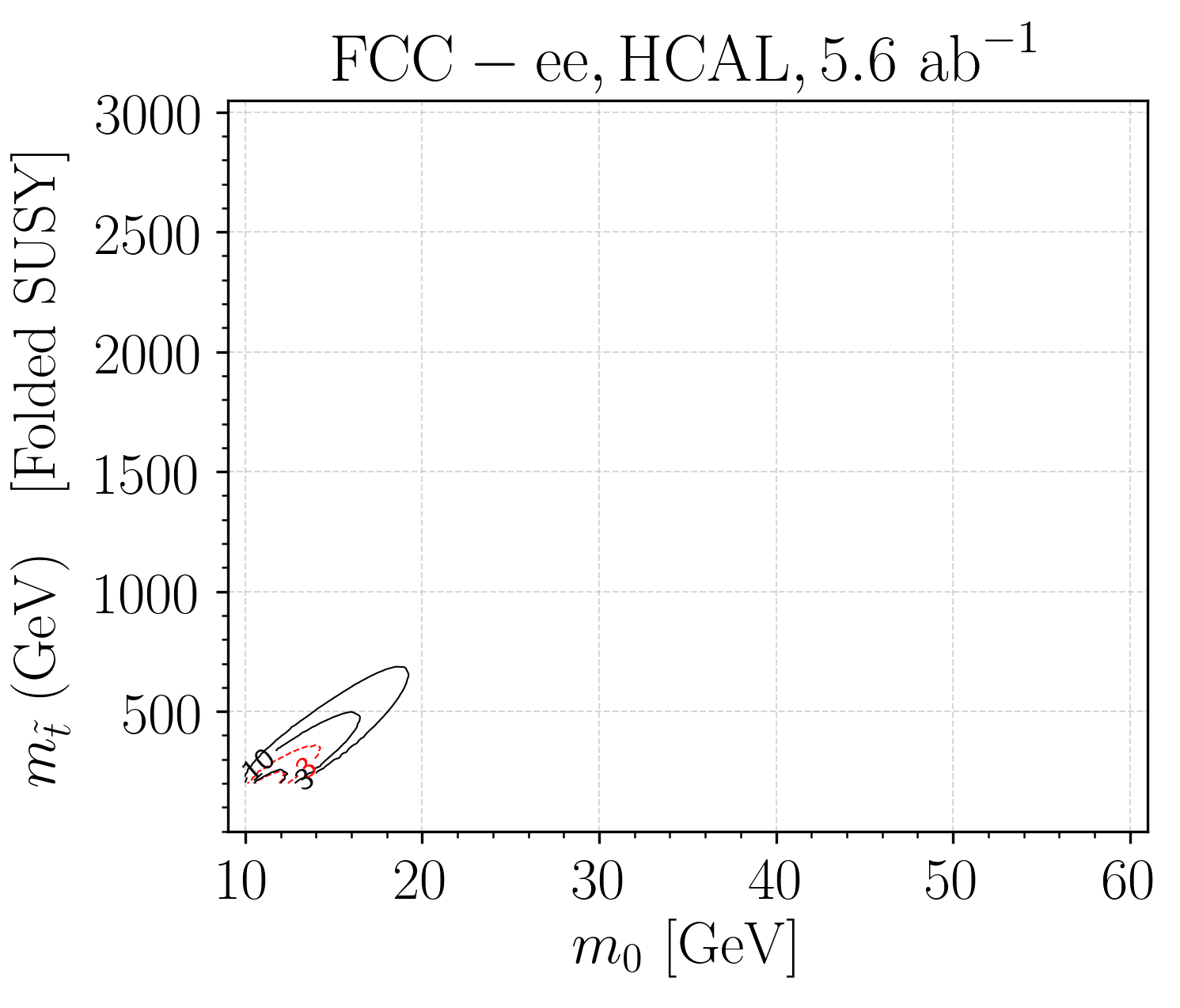}
	\includegraphics[width=0.4\textwidth]{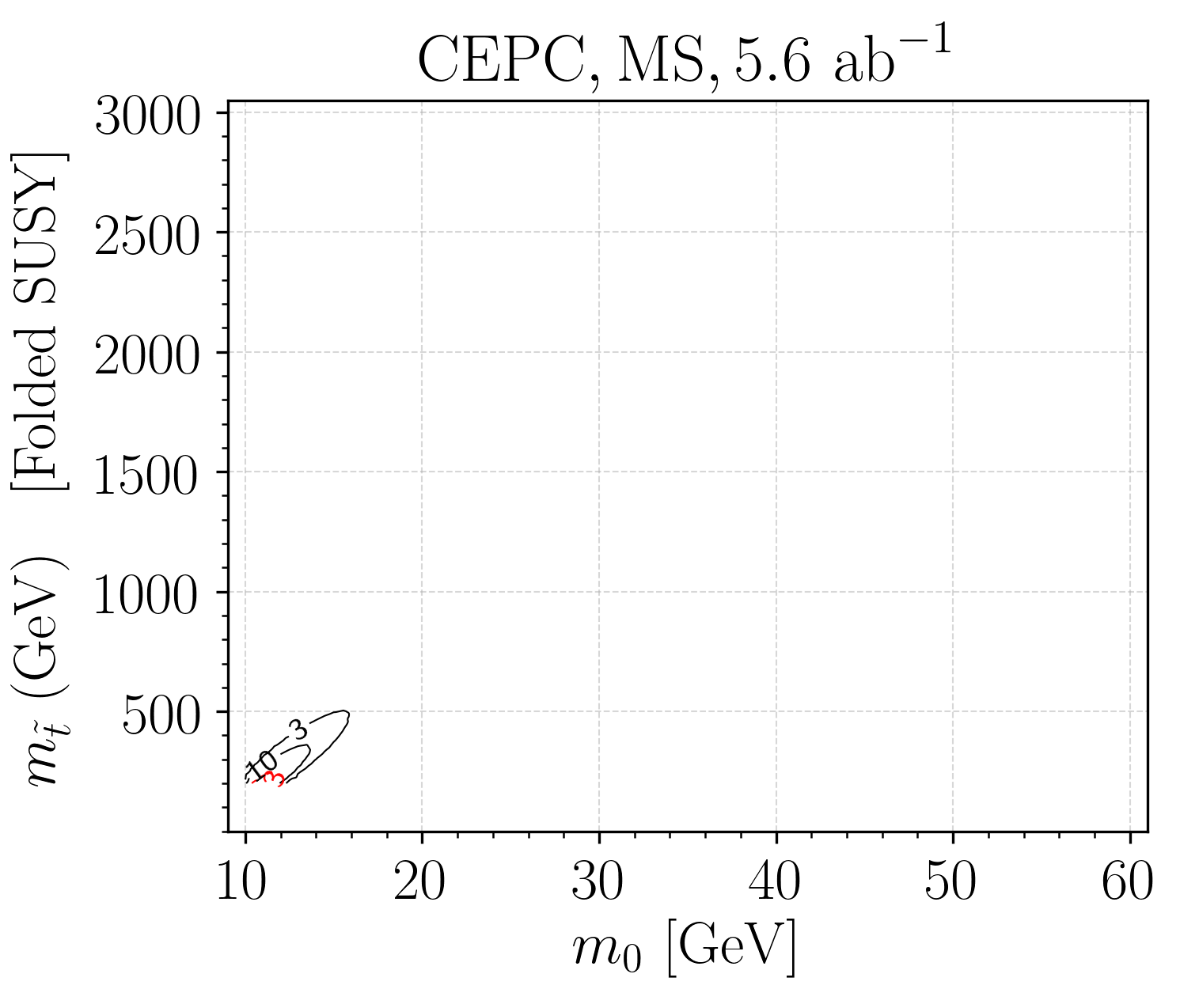}
	\includegraphics[width=0.4\textwidth]{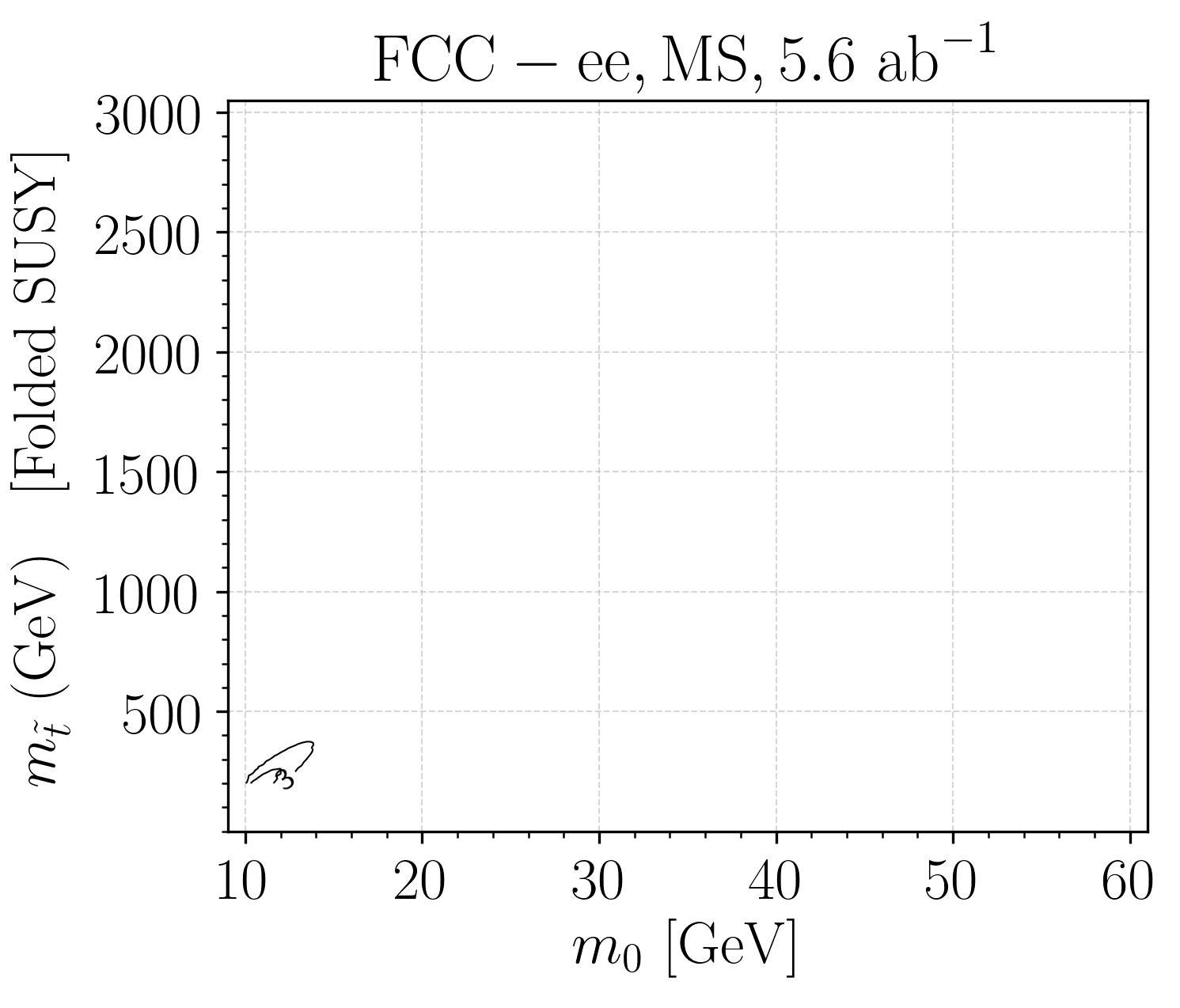}
	\caption{Sensitivity reaches of $\text{log}_{10}(N_{\text{signal}})$
          at the CEPC and FCC-ee for the Folded SUSY model.
          The black(red) curves correspond to
          $\kappa=\kappa_{\text{max}}$ ($\kappa=\kappa_{\text{min}}$)}.
	\label{fig:sensitivity3}
\end{figure}

\begin{figure}
	\includegraphics[width=0.4\textwidth]{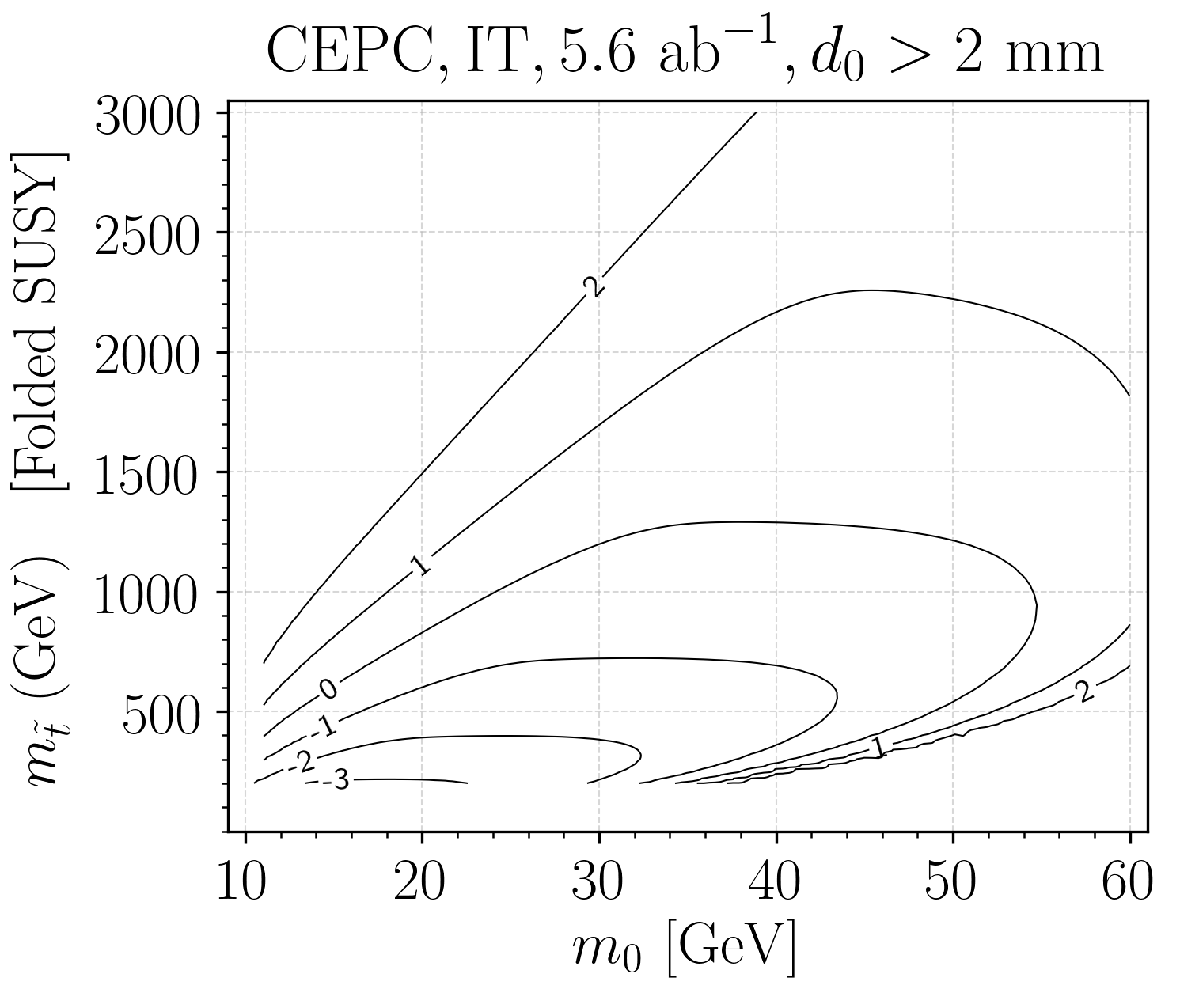}
	\includegraphics[width=0.4\textwidth]{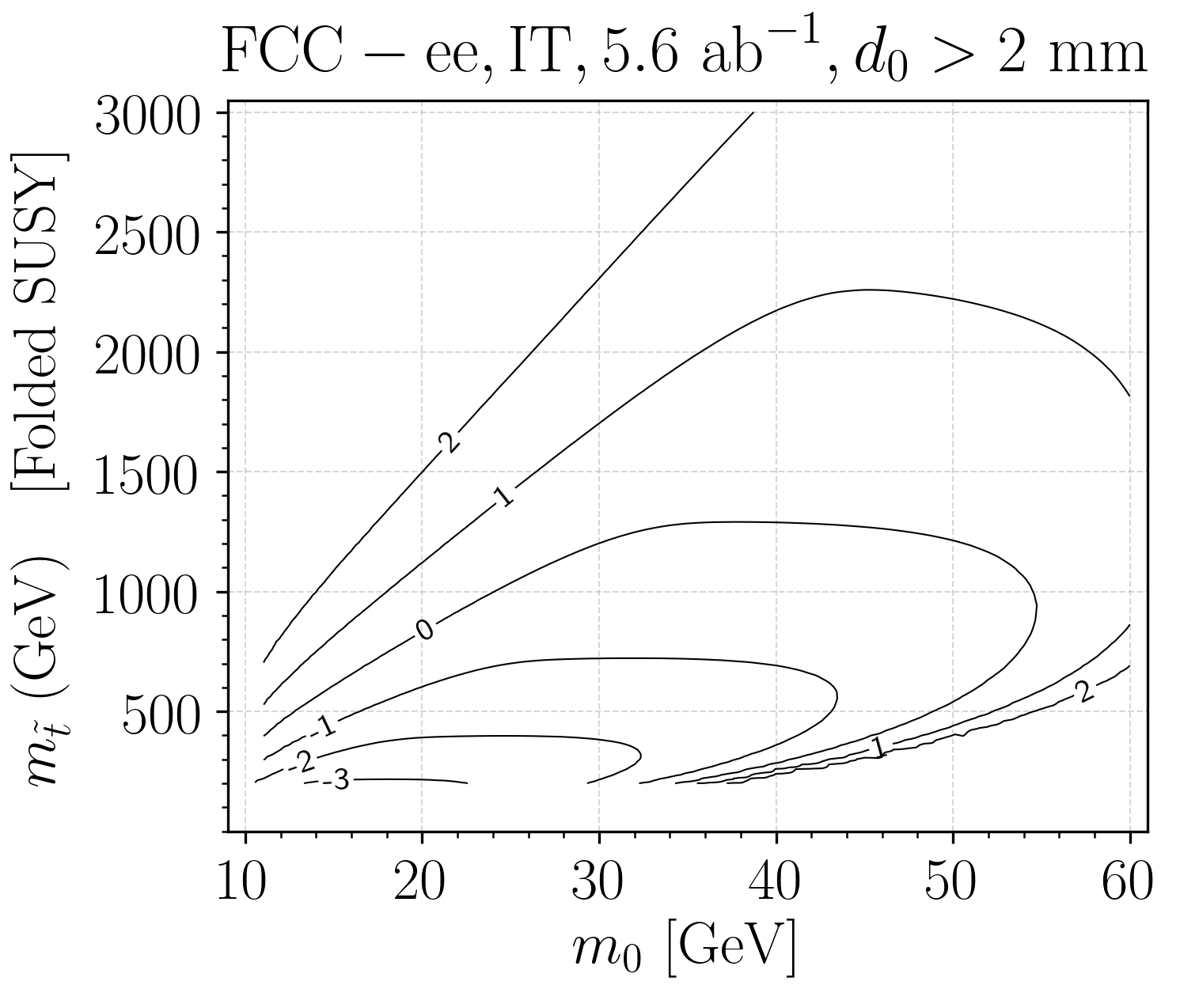}
	\includegraphics[width=0.4\textwidth]{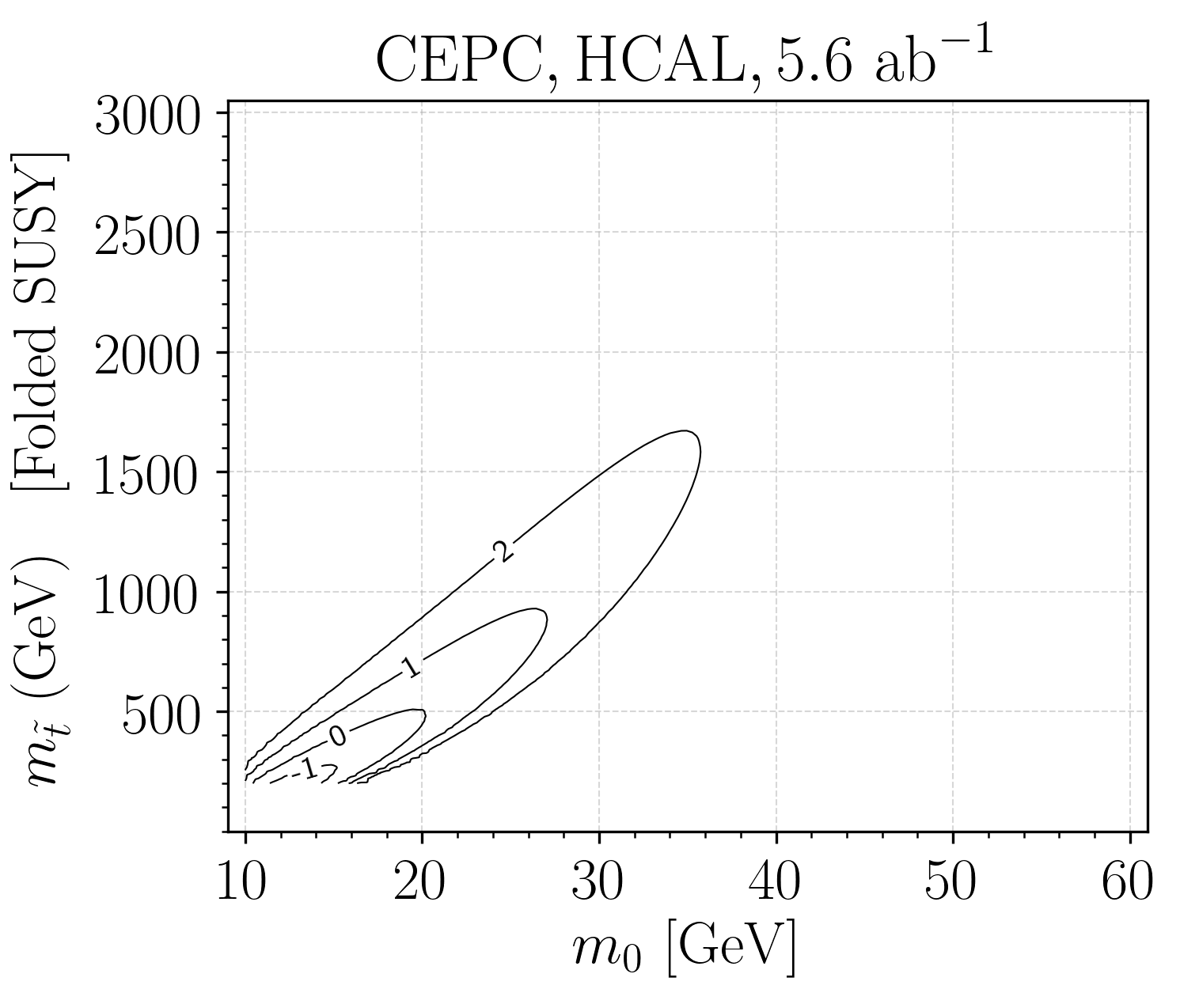}
	\includegraphics[width=0.4\textwidth]{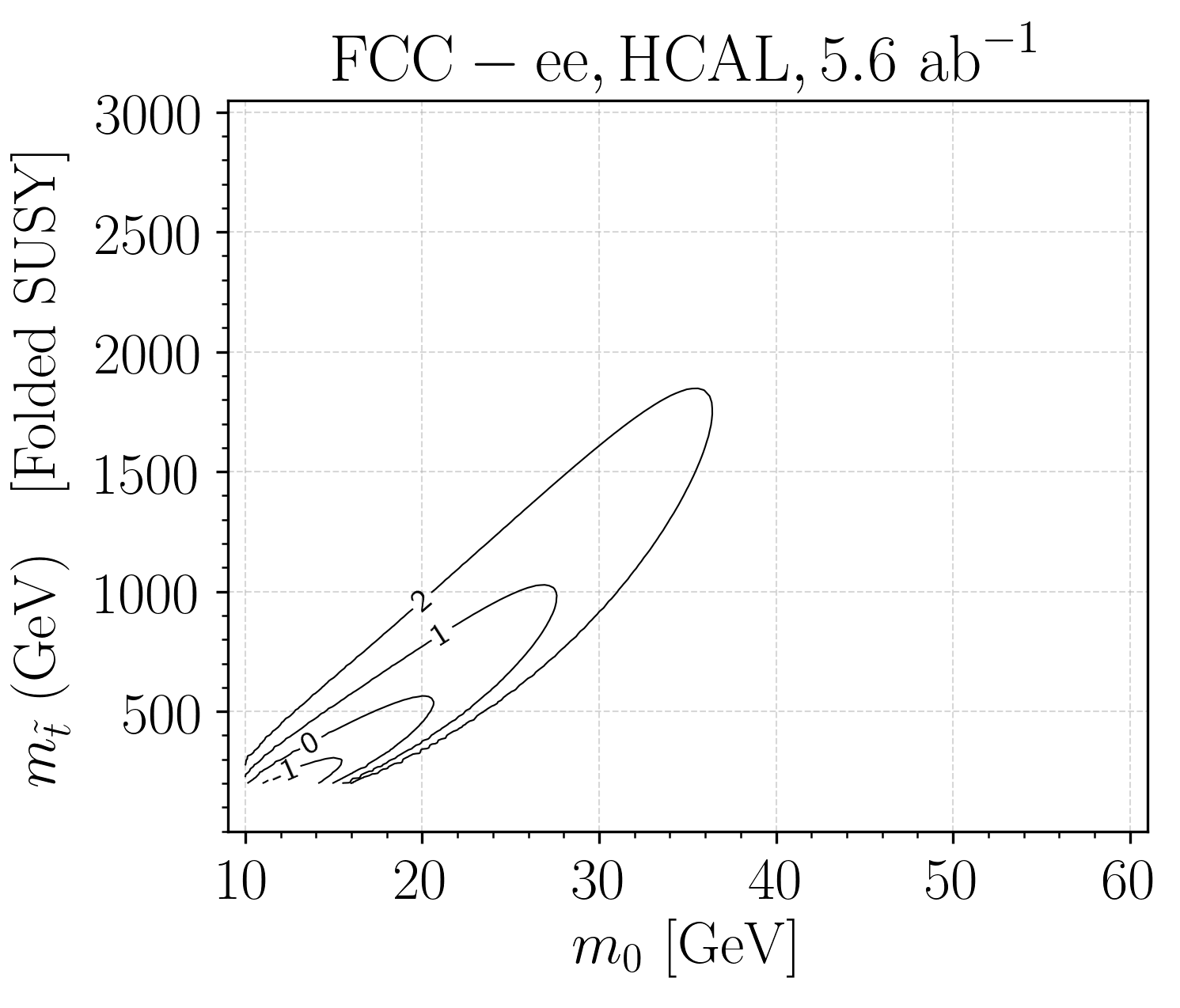}
	\includegraphics[width=0.4\textwidth]{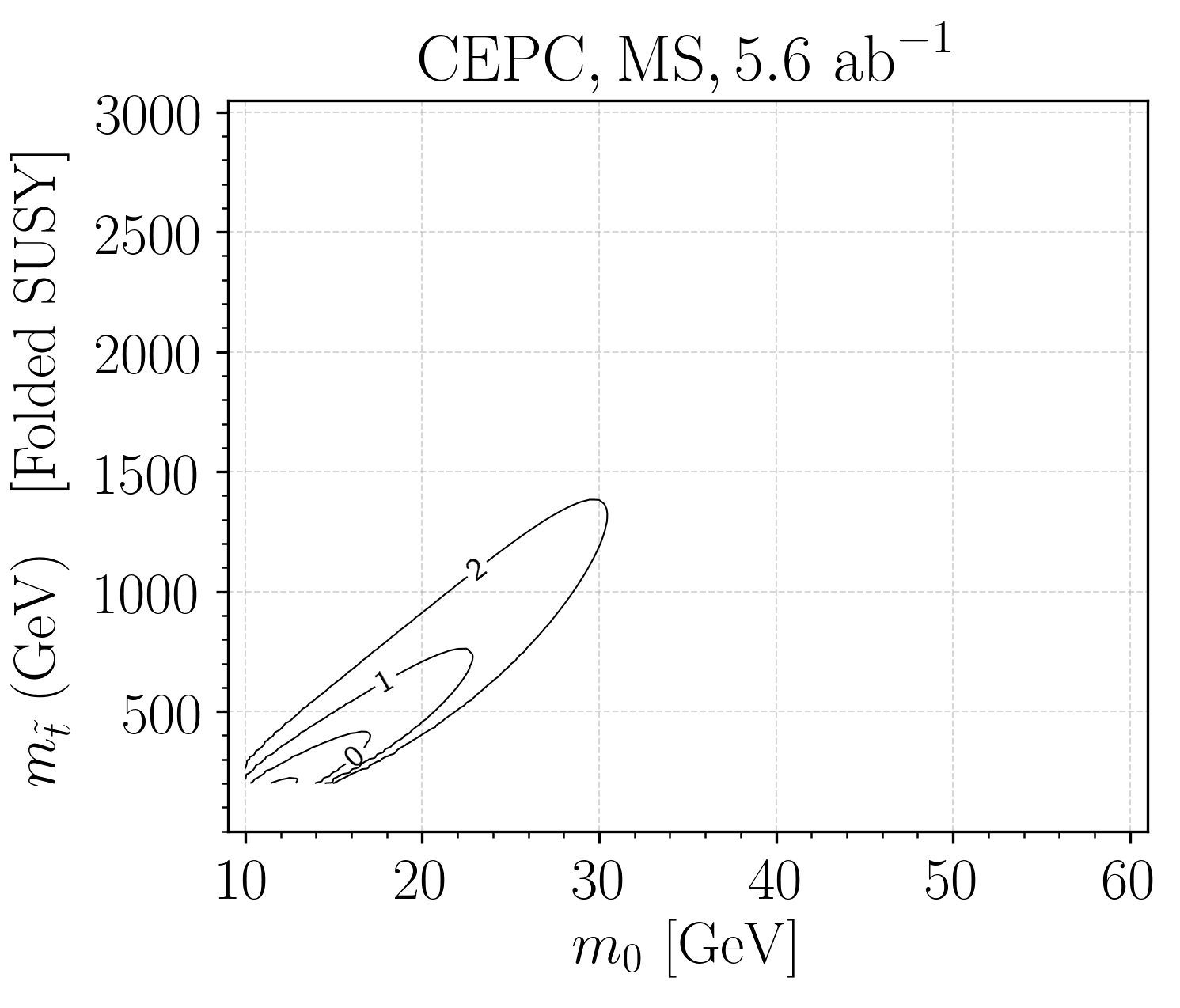}
	\includegraphics[width=0.4\textwidth]{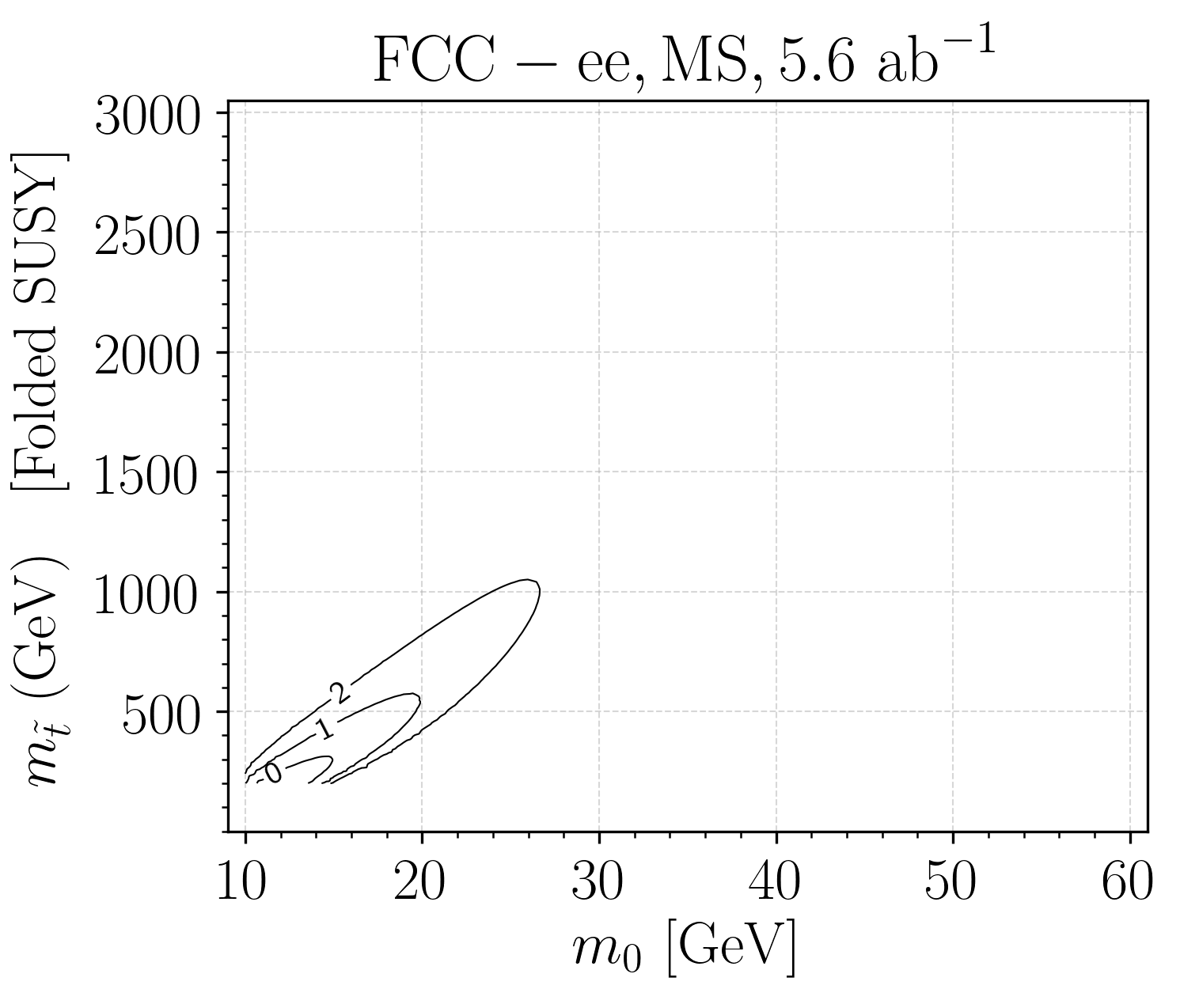}
	\caption{Sensitivity reaches of $\text{log}_{10}(\kappa)$ at the
          CEPC and FCC-ee for the Folded SUSY model for $N_{\text{signal}}=10$.}
	\label{fig:sensitivity4}
\end{figure}

We now present the results for the mirror glueball case. We focus on
their decay into a pair of $b-$jets as this channel has the largest
decay branching ratio for $0^{++}$ in the mass range under consideration,
and consequently we perform sensitivity estimation for all
of the IT, HCAL, and MS.
  There are two major backgrounds under our consideration, i.e.
  $e^- e^+ \to Z Z \to (\ell^+ \ell^-, jj) (b\bar b)$ and
  $e^- e^+ \to Z h \to (\ell^+ \ell^-, jj) (b\bar b)$, in which the
  $b\bar b$ pair comes from prompt $Z$ or $h$ decay. We make use of all
  visible decays ($\ell^+ \ell^-, q\bar q$) of the $Z-$boson, which
  corresponds to a branching ratio of $\sim 0.8$. An invariant mass cut
  is imposed on the lepton pair or quark pair to identify the $Z$ boson.
  It is followed by a recoil-mass cut, defined by
  \[
  M_{\rm recoil}^2  = s - 2 \sqrt{s} ( E_{\ell^+} + E_{\ell^-} ) + M_{\ell\ell}^2\,;
  \qquad
  120 \;{\rm GeV} < M_{\rm recoil} < 150 \;{\rm GeV},
  \]
  in order to remove the $ZZ$ background. It was shown in
  Ref.~\cite{Craig:2015pha}  that such a cut can reduce the $ZZ$
  background down to less than $10^{-2}$.
  In order to remove the prompt
  $h/Z\to b\bar b$ decay and the SM bottomonium background
  events of displaced vertices taking place in the IT,
  we further impose an invariant-mass cut on the
  $b\bar b$ pair: $10 \,{\rm GeV} < M_{b\bar{b}} < 80$ GeV, which has no
  effect on the signal events.
Furthermore, we require the transverse impact parameter $d_0>2$ mm for
both $b$-jets stemming from any secondary vertex in the IT, so as to
make sure that the corresponding vertex is a displaced one and can be
reconstructed. This would cut away some sensitivity at the ultra-short
decay length regime (corresponding to the lower right corner in the
shown plane $m_{\tilde{t}}$ vs. $m_0$).  Similar to the previous model
we have assumed that for the HCAL and MS the SM background is
negligible.

We present our sensitivity estimates in two ways. In
Fig.~\ref{fig:sensitivity3}, we make plots of contour curves for
$N_{\text{signal}}$ denoting the
number of signal events. The selected isocurves are for $ N_{\text{signal}}= 3, 10, 100$.
The left and right columns correspond to the CEPC and FCC-ee, respectively. The black (red) curves are for $\kappa_{\text{max}}$ ($\kappa_{\text{min}}$). In Ref.~\cite{Alipour-Fard:2018lsf}, the limits for the same model were also
shown for the CEPC/FCC-ee with $N_h=1.1\times 10^{6}$. Compared to the results therein, our estimates for the IT (with 3 signal events or 95\% C.L. with 0 background) are more optimistic with the maximal potential reach of $m_{\tilde{t}}$ roughly 2 times better. This is largely due to the fact that we include both $l^- l^+$ and $jj$ decay channels of the $Z-$bosons while Ref.~\cite{Alipour-Fard:2018lsf} considered only leptonic $Z$ decays. Our sensitivity reach in $m_0$ is however smaller, because the requirement on the transverse impact parameter cuts away most signal events for large decay width.
Compared to the IT, the HCAL and MS may have smaller sensitivity
but still useful coverage in the parameter space.

In Fig.~\ref{fig:sensitivity4} we present another set of plots with
contour curves for $\log_{10}(\kappa)$. The selected values of $\log_{10} (\kappa) = -3,\,-2,\, -1,\,0,\,1,\,2$. We may compare our results with those for the HL-LHC with $\sqrt{s}=14$ TeV and 3 ab$^{-1}$ integrated luminosity presented in Ref.~\cite{Curtin:2015fna}. Our IT search may probe $m_{\tilde{t}}$ roughly 1.5 times better while our HCAL and MS limits are similar or slightly worse.

Note that we only show our results for the folded SUSY model. For the
other neutral-naturalness models, as the parameter $y^2/M^2$ is also inversely proportional to the square of the new scale, one can easily obtain the corresponding sensitivity estimates by simple rescaling.

\section{Conclusions}\label{sec:conclusions}

We have investigated the potential of the future
$e^- e^+$ colliders operated as the Higgs factories,
with the profiles of CEPC and FCC-ee as examples, in detecting the
long-lived particles predicted by a number of models beyond the SM.
We have employed two representative models, i.e. the hidden scalar model and
folded SUSY model, which feature two distinct mass ranges, including
sub-GeV and $O(10)$ GeV, respectively.

The decay of the sub-GeV scalar boson gives rise to a pair of
collimated muons or pions, which provides a distinctive signature
against possible SM backgrounds. As a result of a much
larger decay branching ratio into the pion pair the sensitivity
reach at CEPC and FCC-ee can be substantially better than the LHC.

The decay of relatively heavier mirror glueballs of mass $O(10)$ GeV
leads to a pair of $b-$jets with a clean secondary vertex. With
a series of selection cuts all possible SM backgrounds can be
rejected while the signal events remain largely unaffected. By including both the leptonic and hadronic decays of the $Z-$bosons, the sensitivity reach at the CEPC and FCC-ee is about a few
times better than previous studies. 

We offer a few more comments as follows.
\begin{enumerate}
\item
  In this study, we have assumed that with the selection cuts, such as
  collimated muon or pion pairs for the sub-GeV scalar boson and $b-$jet
  pairs with a secondary vertex and large invariant mass
  for the folded SUSY model, most SM backgrounds can be eliminated.
 
\item
  In principle, one can also study the sensitivity reach for the ILC with the
  proposed geometries. However, the designed luminosity is relatively low so that one expects only fewer than 400k Higgs bosons to be produced at $\sqrt{s}=250$ or 500 GeV after a luminosity upgrade \cite{Borzumati:2014zxa}, which is still below the expectation at the CEPC/FCC-ee.

\item
  Another interesting mass range $O(1-10)$ GeV is worth studying, because
  the decay of such a LLP may suffer from the SM backgrounds of $B$-hadrons.
  
\item
  The Folded SUSY model is just an example of the neutral-naturalness models.
  Results for the other similar models can be easily obtained by rescaling
  the parameter $y^2/M^2$.
  
\end{enumerate}

\section*{Acknowledgment}
We thank Nathaniel Craig, Chih-Ting Lu, Yue-Lin Sming Tsai, Michael Williams, and Jos\'e Francisco Zurita for useful discussions.  Z.~S.~W. is supported
by the Ministry of Science, ICT \& Future Planning of Korea, the
Pohang City Government, and the Gyeongsangbuk-do Provincial Government
through the Young Scientist Training Asia-Pacific Economic Cooperation
program of APCTP, and thanks the National Center for Theoretical
Sciences for hospitality where part of this work was conducted.
K.~C. is supported by the MoST of Taiwan under grant
no. 107-2112-M-007-029-MY3.

\bibliographystyle{JHEP}
%\bibliography{llp}

\providecommand{\href}[2]{#2}\begingroup\raggedright\endgroup

\end{document}